\newcommand{\dmunit}{\mathrm{pc\,cm^{-3}}}

\newcommand{\msun}{\mathrm{M}_\odot}
\newcommand{\rsun}{\mathrm{R}_\odot}

\newcommand{\Pb}{P_{\rm b}}

\newcommand{\Mp}{M_{\rm p}}
\newcommand{\Mc}{M_{\rm c}}

\documentclass{aa}

\usepackage{orcidlink}
\usepackage{hyperref}

\hypersetup{hidelinks}

\usepackage{graphicx}

\usepackage{txfonts}

\usepackage{booktabs}

\usepackage{array}
\usepackage{makecell}
\usepackage{placeins}
\usepackage{capt-of}
\usepackage{siunitx}

\begin{document}

\title{The first MeerKAT S-band globular cluster pulsar survey}

\author{
Rouhin Nag \orcidlink{0009-0005-6754-0655}\inst{1,2}
\and
Marta Burgay\orcidlink{0000-0002-8265-4344}\inst{2}
\and
Alessandro Ridolfi\orcidlink{0000-0001-6762-2638}\inst{3,2}
\and
Federico~Abbate\inst{2}
\and
Andrea Possenti\inst{2}
\and
Paulo C. C. Freire\orcidlink{0000-0003-1307-9435}\inst{4}
\and
Scott~M.~Ransom\inst{5}
\and
Miquel Colom i Bernadich\inst{2,4}
\and
Michael Kramer\orcidlink{0000-0002-4175-2271}\inst{4,6}
\and
Benjamin W. Stappers\inst{6}
\and
Ewan~D.~Barr\inst{4}
\and
Rene P. Breton\orcidlink{0000-0001-8522-4983}\inst{6}
\and
Vivek Venkatraman Krishnan\inst{4}
\and
Prajwal~V.~Padmanabh\inst{7,8}
\and
Weiwei Chen\inst{4}
\and
David~J.~Champion\inst{4}
\and
Maciej Serylak\orcidlink{0000-0002-6670-652X}\inst{9}
\and
Alessandro Corongiu\inst{2}
\and
Mario Cadelano\inst{10,11}
\and
Vishnu Balakrishnan\inst{12}
\and
Arunima Dutta\inst{4}
\and
Dhanraj Risbud\inst{4,3}
}
\institute{
Department of Physics, University of Cagliari, Cittadella Universitaria di Monserrato, 09042 Monserrato, Italy
\and
INAF -- Astronomical Observatory of Cagliari (OAC), Via della Scienza 5, 09047 Selargius, Italy
\and
Fakultät für Physik, Universität Bielefeld, Postfach 100131, D-33501 Bielefeld, Germany
\and
Max-Planck-Institut für Radioastronomie, Auf dem Hügel 69, D-53121 Bonn, Germany
\and
National Radio Astronomy Observatory, 520 Edgemont Rd., Charlottesville, VA, 22903, USA
\and
Jodrell Bank Centre for Astrophysics, Department of Physics and Astronomy, The University of Manchester, Manchester, M13 9PL, UK
\and
Max Planck Institute for Gravitational Physics (Albert Einstein Institute), D-30167 Hannover, Germany
\and
Leibniz Universit\"{a}t Hannover, D-30167 Hannover, Germany
\and
SKA Observatory, Jodrell Bank, Lower Withington, Macclesfield, Cheshire, SK11 9FT, UK
\and
Dipartimento di Fisica e Astronomia, Università degli Studi di Bologna, Via Gobetti 93/2, I-40129 Bologna, Italy
\and
INAF, Osservatorio di Astrofisica e Scienza dello Spazio di Bologna, Via Gobetti 93/3, I-40129 Bologna, Italy
\and
Center for Astrophysics | Harvard \& Smithsonian, Cambridge, MA 02138-1516, USA
}

\abstract
{Globular clusters are efficient factories of recycled pulsars, but searches towards high dispersion-measure (DM) clusters can be strongly limited around 1\,GHz---which is where most pulsar surveys are typically conducted---due to dispersive smearing and interstellar scattering. Observations at higher radio frequencies provide a complementary route to mitigate these effects and improve sensitivity.}
{We present the first results from a MeerKAT S-band ($\nu\simeq2.4$\,GHz) pulsar survey of 14 globular clusters. Our aims were to discover new pulsars, re-detect known pulsars with a uniform high-frequency dataset, and assess practical S-band sensitivity and DM-search requirements for future globular cluster pulsar surveys.}
{High time and frequency resolution S-band observations were searched in the Fourier domain using segmented acceleration and jerk techniques to maintain sensitivity to pulsars in compact binaries, followed by candidate identification and folding. For selected detections and non-detections, we additionally folded the S-band data together with archival MeerKAT L-band observations using available timing ephemerides to derive flux-density measurements and spectral-index constraints.}
{We re-detected 39 previously known pulsars across 14 clusters and discovered four new millisecond pulsars in Glimpse-C01 (J1848$-$0129C, D, E, and F). Multi-epoch follow-up enabled preliminary Keplerian orbital fits for two of the new systems based on their observed barycentric spin-period modulation. J1848$-$0129C is an eclipsing millisecond pulsar with $\Pb \sim$ 5 d, placing it among the long-period eclipsing systems known as `huntsman' binaries. J1848$-$0129D is in a $\Pb \sim 3.4\, \rm d$ nearly circular orbit with a $\sim 1\, \msun$ massive white dwarf companion. Two Glimpse-C01 pulsars exhibit substantial DM offsets relative to the average cluster value. Comparison with other Galactic globular clusters showed that large intracluster DM spreads become increasingly common at high foreground DMs, implying that searches confined to narrow DM windows may be sub-optimal for high-DM clusters. From detections and non-detections, we inferred a practical single-epoch S-band tied-array detectability scale of $\sim$10--20~$\mu$Jy, which serves as an operational benchmark rather than a strict completeness limit.}
{MeerKAT S-band observations are highly effective for probing globular clusters affected by strong dispersion and scattering, enabling both reliable re-detections and the discovery of new millisecond pulsars in high-DM environments such as Glimpse-C01. The results of our work highlight the importance of high-frequency surveys for expanding the known pulsar population in obscured clusters.}


   \maketitle

\section{Introduction}

Globular clusters (GCs) are among the most productive environments for pulsar searches. These dense, gravitationally bound systems contain $\sim10^{4}$--$10^{6}$ stars packed into compact volumes, creating ideal conditions for frequent dynamical encounters. Such interactions can cause the exchange of companions, harden binaries, and recycle neutron stars through the transfer of mass and angular momentum from a companion star, efficiently producing binary millisecond pulsars (MSPs) that are otherwise comparatively rare in the Galactic field \citep{Alpar1982,Bhattacharya1991}. For this reason, GCs are attractive targets for deep observations, as a single long pointing is likely to host multiple pulsars and can build the signal-to-noise ratio for intrinsically faint systems or those affected by scattering and eclipses. At the same time, long dwell times can be detrimental to the detection of pulsars in compact binaries, since orbital acceleration and higher-order motion can smear the signal over extended integrations if not properly accounted for.

The scientific motivation for GC pulsar surveys is correspondingly broad. Recycled pulsars in compact binaries provide precision laboratories for tests of relativistic gravity \citep[e.g.][]{1982ApJ...253..908T,2021PhRvX..11d1050K}, constraints on the neutron-star equation of state \citep{2016ARA&A..54..401O}, and the study of exotic binaries formed through dynamical interactions in dense stellar systems \citep[e.g.][]{2024Sci...383..275B}. Ensembles of stable millisecond pulsars contribute to pulsar timing array experiments targeting nanohertz gravitational waves \citep[e.g.][and references therein]{Tiburzi2018,Mingarelli2026}, while GC pulsars, by virtue of their common distance and shared environment, serve as sensitive probes of cluster gravitational potentials and the intervening interstellar medium \citep{2017MNRAS.471..857F}. These drivers have motivated sustained pulsar-search campaigns over several decades with facilities including Murriyang, the radio telescope at Parkes (New South Wales, Australia), the Green Bank Telescope (GBT; West Virginia, USA), as well as with more recently upgraded instrumentation and new high-sensitivity facilities such as the upgraded Giant Metrewave Radio Telescope (GMRT; India) and the Five-hundred-metre Aperture Spherical Telescope (FAST; Guizhou, China).

MeerKAT is a 64-element interferometer array located in the Northern Cape province of South Africa, and its dishes will be incorporated in the intermediate-frequency telescope of the SKA Observatory \citep[SKA-Mid;][]{SKA-Mid2022}. In recent years it has become a cornerstone facility for pulsar science in the southern hemisphere. The combination of collecting area, modern receivers, and wide instantaneous bandwidth has made it a highly effective instrument for deep pulsar searches and follow-up timing observations, including in GCs (see \citealt{Bailes2020}). The impact of MeerKAT on GC pulsar studies is already evident through the Transients and Pulsars with MeerKAT \citep[TRAPUM;][]{StappersKramer2016} and MeerTime \citep{Bailes2018} programmes, the two pulsar-focused Large Survey Projects carried out with MeerKAT over the past five years. Together, these projects have delivered a substantial fraction of recent GC pulsar discoveries and continue to expand the accessible parameter space for faint systems and compact binaries \citep{Ridolfi2021,Ridolfi2022,Abbate_2022}.

Most GC pulsar surveys to date have been conducted at frequencies near 1–1.4\,GHz, where pulsar flux densities are typically higher compared to higher frequency observations and receiver performance has historically been optimised. However, for distant clusters with large foreground dispersion measures (DMs), pulse detectability at such frequencies can be significantly limited by dispersive smearing and, more critically, by multi-path scattering in the interstellar medium. The scattering timescale decreases steeply with observing frequency, approximately as $\tau_{\mathrm{sc}} \propto \nu^{-4}$ \citep{2004ApJ...605..759B}, such that millisecond pulsars in high-DM systems may suffer substantial pulse broadening at the L-band. Higher frequency searches have therefore been explored with single-dish facilities including Effelsberg, the GBT, and Parkes in targeted studies of heavily scattered or high-DM pulsars, leading to successful detections in otherwise challenging systems \citep{Ransom2005,2006MNRAS.373L...6J}. Over the past two decades, several GCs have also been observed at 2\,GHz with the GBT in dedicated campaigns \citep[e.g.][]{2011ApJ...730L..11L,2015ApJ...807L..23D}, demonstrating the utility of higher observing frequencies for mitigating dispersive smearing and scattering in dense Galactic-plane sightlines.

Systematic, uniformly processed multi-cluster surveys above 2\,GHz remain comparatively uncommon. MeerKAT’s large collecting area, coherent tied-array beamforming capability, and wide instantaneous bandwidth provide a complementary high-sensitivity platform for conducting systematic S-band ($\nu \simeq 2.4$\,GHz) searches across a broader cluster sample. The work presented here reports the preliminary results of a MeerKAT S-band GC pulsar survey of 14 clusters observed during 2023--2024 as part of TRAPUM. We present the outcomes of the search and folding campaign, including the discovery of four MSPs in the GC Glimpse-C01 (see Fig.~\ref{fig:phase_time}) and the re-detection of 39 previously known pulsars across the full sample (see Table~\ref{tab:gc_observations} and Sect.~\ref{sub_re_det}). Beyond reporting detections, our aim was also to quantify practical S-band performance for future studies. To that end, we (i) estimated S-band flux densities for the detected pulsars, (ii) placed these measurements in context by comparing a subset to representative (archival) MeerKAT L-band flux-density values for the same objects (see Tables~\ref{tab:flux_simple} and~\ref{tab:ter5_flux_density}), and (iii) derived illustrative spectral-index constraints for selected non-detections using S-band upper limits (see Sect.~\ref{sec:sband_nondetections}).

\section{Target selection and observations}

For this survey, 14 GCs were observed with the S1-band of the MeerKAT S-band receiver system, covering a frequency range of 1750–3500\,MHz, with an effective usable bandwidth of $\Delta\nu \simeq 875$\,MHz centred near 2406.25\,MHz over a one-year period starting in March 2023. These GCs are listed in Table~\ref{tab:gc_observations}. All targets host at least one previously known pulsar and were previously observed with MeerKAT at L-band as part of the TRAPUM survey \citep{StappersKramer2016,Ridolfi2021,Ridolfi2022,Abbate_2022,Douglas_2022,Vleeschower_2022,Chen_2023} except NGC6316, Glimpse-C01, NGC5986, NGC6539 and NGC6712.

Several clusters in our sample have previously been searched at S-band or comparable frequencies with other facilities, including Terzan~5, NGC~6440, NGC~6441, $\omega$~Centauri, and M28 \citep[e.g.][]{Ransom2005,Freire2008,2011ApJ...730L..11L,2012ApJ...745..109L,Dai2020}. NGC~6316 has also been targeted in a recent GBT S-band study \citep{Bhakta2026NGC6316}. For the remaining clusters in our sample, no dedicated S-band pulsar searches have, to the best of our knowledge, been performed.

Targets were selected using complementary criteria intended to probe the strengths and limitations of MeerKAT S-band pulsar searching in GCs. Clusters with high DMs were prioritised because higher observing frequencies reduce dispersive smearing and multipath scattering, thereby preserving narrower pulse profiles and higher effective signal-to-noise ratios for fast-spinning MSPs. This does not directly change the range of orbital accelerations probed by the search, which is set by the integration length and the acceleration/jerk search parameters, but it improves the detectability of compact binary pulsars by making their pulsed signals less vulnerable to propagation-induced broadening, especially when the data are searched in shorter sub-integrations. In parallel, several clusters with already rich pulsar populations were re-observed to maximise the likelihood of re-detections, to confirm candidates, and to provide a consistent dataset for assessing S-band sensitivity for future surveys.

Each observation used a four-beam Pulsar Timing User Supplied Equipment (PTUSE) tied-array configuration \citep{Chen2021JAI}. For every pointing, one full-array beam (typically 56--64 antennas, subject to availability) and one `1 km' core beam (typically 38--44 inner-core antennas) were centred on the nominal cluster position (see Fig.~\ref{fig:beammaps}). The two beam configurations, typically a few arcseconds across for the full array and tens of arcseconds across for the 1\,km core, sample different regions of parameter space: the full-array beam provides higher sensitivity over a narrower field of view, while the 1\,km core beam offers lower sensitivity but larger on-sky coverage, improving robustness to positional uncertainties within the cluster. As illustrated in Fig.~\ref{fig:beammaps}, a single 1\,km beam covers the core region of several compact clusters but only part of the core in the more extended systems.

Where sufficiently precise timing positions (i.e. at least a few arcseconds) were available for previously known pulsars (e.g. NGC~6624, NGC~5986, NGC~6539, NGC~6544), the remaining two beams (again, one full-array and one 1\,km beam) were pointed at the known pulsar location(s). In these cases, re-detections were obtained through blind searches as well as coherent folding using published timing ephemerides, with the aim of confirming detectability, assessing signal stability, and, in selected cases, verifying the consistency of existing timing solutions (e.g. NGC~6712A; \citealt{Yan2021NGC6712}) or candidates reported in earlier surveys (e.g. Glimpse-C01A; \citealt{McCarver2024GlimpseC01}).

\section{Data analysis}

\subsection{Data format for the S-band survey}

The pulsar search was performed on search-mode PSRFITS data products (see \citealt{Hotan2004}) generated as part of the standard MeerKAT pulsar processing chain. The data consist of total-intensity PSRFITS files, formed by summing the two recorded linear polarisations, and are channelised into 256 frequency channels spanning the full S-band bandwidth. The data are dedispersed coherently at the nominal cluster DM used for the search, and have a final time resolution appropriate for millisecond-pulsar detection corresponding to the cluster DM's (75 or 150 $\mu$s; see Table~\ref{tab:gc_observations}).

\subsection{Search operation}

All 14 GCs in this survey were searched using \texttt{PULSAR MINER} v2.0-beta \footnote{\url{https://github.com/alex88ridolfi/PULSAR_MINER.git}}, an automated pulsar-search pipeline built on the \texttt{PRESTO} software suite \citep{2011ascl.soft07017R}. All the PSRFITS files were first processed to further mitigate radio-frequency interference (RFI). An RFI mask was generated for each observation using the \texttt{rfifind} routine. Rather than adopting a single set of default parameters, we manually adjusted the time and frequency integration scales and the channel- and interval-rejection thresholds for each dataset. The resulting mask statistics and diagnostic plots were then inspected to ensure that any persistent RFI was removed.

The MeerKAT S-band data were generally clean, and in most observations the RFI excision performed with \texttt{rfifind} was sufficient for the subsequent search. For the small number of observations affected by narrow periodic RFI, we generated a zero-DM time series using \texttt{prepdata} and used it to identify obvious instrumental or anthropogenic features. These features were added to an observation-specific birdie list and the corresponding Fourier frequencies were masked before the pulsar search.

As in standard pulsar search pipelines, a de-dispersion plan tailored to each observation was then constructed using the \texttt{DDplan.py} utility in \texttt{PRESTO}. This procedure automatically determines the trial-DM spacing and any required down-sampling from the instrumental setup, sampling interval, coherent de-dispersion DM, and requested DM range, balancing the smearing caused by finite DM spacing against the effective time resolution and intra-channel smearing. Because the instrumental configuration was largely uniform across the survey, the generated plans differed mainly in the nominal DM, searched DM range, and native sampling interval. For this survey, the overall DM search range was chosen to extend to within $\pm 20\%$ of the nominal cluster DM. This was a conservative choice to remain sensitive to certain outlier pulsars that could lie along lines of sight with notable high electron density structures especially for high DM clusters like Glimpse-C01.

As a representative check on the resulting time resolution, we examined the de-dispersion plans for M22, NGC~6440, and Glimpse-C01, which span nominal DMs of approximately 91, 223, and 491~pc~cm$^{-3}$, respectively. The adopted trial-DM spacings were 0.5, 0.3, and 0.5~pc~cm$^{-3}$. For a pulsar located halfway between adjacent trials, these correspond to residual dispersive broadening across the 875-MHz band of order 0.08--0.14~ms. When combined in quadrature with the native sampling time and the smaller instrumental contributions included by \texttt{DDplan.py}, the representative effective resolutions are approximately 0.12~ms for NGC~6440 and 0.2--0.25~ms for M22 and Glimpse-C01. These values should be regarded as approximate, since the usable bandwidth and masked channels vary slightly between observations.

This broadening does not define a sharp minimum detectable spin period, but progressively reduces sensitivity when it becomes comparable to the intrinsic pulse width. For a nominal 10\% duty cycle, it would begin to affect the recovery of very narrow pulsars with periods of order 1--2~ms, while pulsars with periods of several milliseconds remain well resolved unless multipath scattering dominates. Using the empirical DM--scattering relation of \citep{2004ApJ...605..759B}, the expected scattering at 2.4~GHz is negligible for the lower-DM clusters but may reach an order of magnitude of $\sim$1~ms towards the highest-DM sightlines, with substantial line-of-sight scatter. Since most of the measured DM is expected to accumulate through the Galactic foreground, the Galactic interstellar medium is likely to provide the dominant scattering contribution. However, the present observations do not allow the Galactic and intracluster contributions to be separated directly.

To recover pulsars in compact binary systems, the data were searched for periodic signals in the Fourier domain using both acceleration and jerk search techniques. In a binary system, orbital motion induces a line-of-sight acceleration that Doppler-shifts the apparent spin frequency of the pulsar, causing its signal power to drift across multiple Fourier bins over the duration of an observation. For a constant acceleration $a$, the resulting frequency derivative is given by $\dot{f} \simeq (a/c)\,f$, where $f$ is the intrinsic spin frequency and $c$ is the speed of light \citep{Ransom2002}. If uncorrected, this frequency drift reduces sensitivity in standard Fourier searches.

The acceleration search, implemented using \texttt{accelsearch}, assumes that the pulsar experiences a constant line-of-sight acceleration over the integration time and coherently recovers signal power by summing across adjacent Fourier bins corresponding to trial acceleration values \citep{Ransom2002}. In PRESTO, the search extent is parameterised by $z_{\max}$, the maximum number of Fourier bins through which a signal is allowed to drift during the observation. In this work, acceleration searches were performed with $z_{\max}=200$. The jerk search extends this approach by additionally allowing for a linear change in acceleration during the observation, thereby increasing sensitivity to systems for which the constant-acceleration approximation breaks down, such as very compact binaries or observations obtained near orbital phases of rapidly varying acceleration \citep{Andersen2018}. In this case, PRESTO introduces a second parameter, $w_{\max}$, which sets the maximum allowed second-order Fourier-bin drift associated with a constant line-of-sight jerk. Our jerk searches were carried out with $z_{\max}=200$ and $w_{\max}=600$. In both the acceleration and jerk searches, up to 16 harmonics were summed in order to improve sensitivity to narrow pulse profiles.

While acceleration and jerk searches substantially improve sensitivity to binary pulsars, their underlying assumptions remain approximations and can still lead to residual signal smearing over long integrations. To mitigate this effect and to control computational cost, the data were therefore also searched at multiple effective integration lengths \citep{JohnsonKulkarni1991}. For each observation, time series were prepared for 15, 30, and 60-minute sub-integrations, in addition to the full $\sim$2-hour observation. Shorter integrations reduce the accumulated frequency drift and improve detectability for the most compact or highly accelerated systems, at the expense of raw signal-to-noise.

\subsection{Sifting and folding}

At the end of the periodicity search, candidate lists were generated for each time-series chunk and trial DM. Here, a candidate corresponds to a statistically significant peak in the Fourier power spectrum (or in a harmonically summed spectrum) exceeding a predefined detection threshold. Candidates were grouped according to harmonic relationships and sifted using standard criteria, requiring that a signal be significant in at least three adjacent DM trials. Only candidates with a detection significance of $\geq 4\sigma$ were retained (see \citealt{Ransom2002}). Depending on the level of residual RFI, the presence of bright or long-period pulsars, and the integration length, this procedure typically produced $\sim$200--500 candidates per beam.

All surviving candidates were then folded using the \texttt{prepfold} routine, using the detected spin period, DM, and trial acceleration as initial parameters. During folding, these parameters were refined to maximise the $\chi^2$ of the folded pulse profile, as compared to an average unpulsed background. The resulting diagnostic outputs include integrated pulse profiles, time--phase and frequency--phase plots, DM response curves, and frequency--frequency-derivative planes.

The folded candidates were subsequently inspected visually to identify broadband signals that persisted over a significant fraction of the observation, showed a well-defined response peaking at a non-zero DM, and produced a stable pulse profile with a width consistent with the expected effective time resolution and dispersive smearing. Promising candidates were cross-checked against the catalogue of known GC pulsars maintained by the Max Planck Institute for Radio Astronomy (MPIfR) \footnote{\url{https://www3.mpifr-bonn.mpg.de/staff/pfreire/GCpsr.html}}. For each cluster, known pulsars were searched for explicitly, including up to 16 harmonics of the fundamental spin frequency, particularly in systems hosting bright or long-period pulsars.

In addition to the blind search described above, targeted folding was performed for all known pulsars covered by the S-band beams using the best existing timing ephemerides at our disposal. For this purpose, available MeerKAT parameter files were used to fold both the S-band census observations presented in this work and multiple epochs of archival MeerKAT L-band data using \texttt{dspsr}\footnote{\url{https://dspsr.sourceforge.net/}} \citep{dspsr2011}. These coherently folded profiles form the basis of the flux-density measurements and spectral-index estimates discussed later in Sects.~\ref{Flex_density_measurements} and~\ref{sec:sband_nondetections}.

\subsection{Flux density measurements}
\label{sec:Flux_calc}

An important component of this work is the measurement of mean flux densities for a subset of re-detected pulsars, with the aim of providing reference values and empirical sensitivity estimates for future observations with MeerKAT or other telescopes at S-band. Flux-density measurements are restricted to pulsars for which the detections yield sufficient signal-to-noise to permit robust baseline and on-pulse flux estimation. Systems exhibiting strong eclipses or highly variable detectability (e.g. spider binaries) are excluded where the observed signal is not representative of the intrinsic pulsar emission.

Rather than relying on direct calibration with the MeerKAT bright noise diode system, which has been noted to potentially affect correlator--beamformer linearity, we estimate mean flux densities using the standard pulsar radiometer equation applied to folded pulse profiles, following the methodology described by \citet{Gitika2023}. Throughout this work, we express the radiometer equation in terms of the effective system-equivalent flux density (SEFD), defined as $\mathrm{SEFD} = T_{\mathrm{sys}}/G$, where $T_{\mathrm{sys}}$ is the total system temperature and $G$ is the telescope gain. The determination of the effective tied-array SEFD for our S-band observations is described in Appendix~\ref{app:sefd}.

For a folded pulsar observation with integration time $t_{\rm obs}$ and usable bandwidth $\Delta\nu$, the phase-averaged mean flux density $S$ is given by
\begin{equation}
S =
\frac{{\rm S/N} \times \mathrm{SEFD}}
{\sqrt{n_{\rm pol}\,\Delta\nu\,t_{\rm obs}}}
\sqrt{\frac{N_{\rm on}}{N_{\rm off}}} ,
\end{equation}
where ${\rm S/N}$ is the signal-to-noise ratio of the integrated pulse profile, $n_{\rm pol}=2$ is the number of summed orthogonal polarisations, $N_{\rm on}$ is the number of phase bins containing pulsed emission, $N_{\rm bin}$ is the total number of phase bins in the folded pulse profile, and $N_{\rm off} = N_{\rm bin} - N_{\rm on}$ is the number of off-pulse bins used to estimate the baseline noise level. The signal-to-noise ratio is measured directly from the folded profile as
\begin{equation}
{\rm S/N} =
\frac{A}{\sigma_A},
\end{equation}
where $A$ is the integrated pulse amplitude above the baseline and $\sigma_A$ is its associated uncertainty derived from the off-pulse rms (see Appendix~\ref{flux_density_method}).

The on-pulse region was identified following the procedure described by \citet{Gitika2023}. An initial selection of contiguous phase bins exceeding the off-pulse baseline rms by a fixed threshold was performed. The selected region was visually inspected as a quality-control step, to check that the off-pulse window was free from obvious baseline structure, residual RFI, or secondary profile components that had been missed by the thresholding step. The remaining bins were designated as the off-pulse region and used to determine the baseline level and noise properties.

Uncertainties on the derived flux densities are obtained by propagating the uncertainty in the measured S/N and the off-pulse rms through Eq.~(1). Since the SEFD, bandwidth, number of polarisations, and integration time are fixed for a given observation, the dominant source of statistical uncertainty arises from the profile noise estimate.

\section{Results}

\subsection{Re-detections}
\label{sub_re_det}

This survey resulted in the re-detection of 39 pulsars across 14 GCs. Details of all re-detections, including the specific tied-array beams in which each pulsar was detected, are summarised in Table~\ref{tab:gc_observations}. The beam information, together with pulsar parameters such as spin period and dispersion measure obtained from the MPIfR catalogue, provides a practical basis for subsequent timing analysis using the S-band observation data archived in the long-term storage facilities of the Astronomical Observatory of Cagliari.

\subsection{New discoveries and other search results from Glimpse-C01}
\label{discoveries}

\begin{figure}
    \centering
    \includegraphics[width=0.68\linewidth]{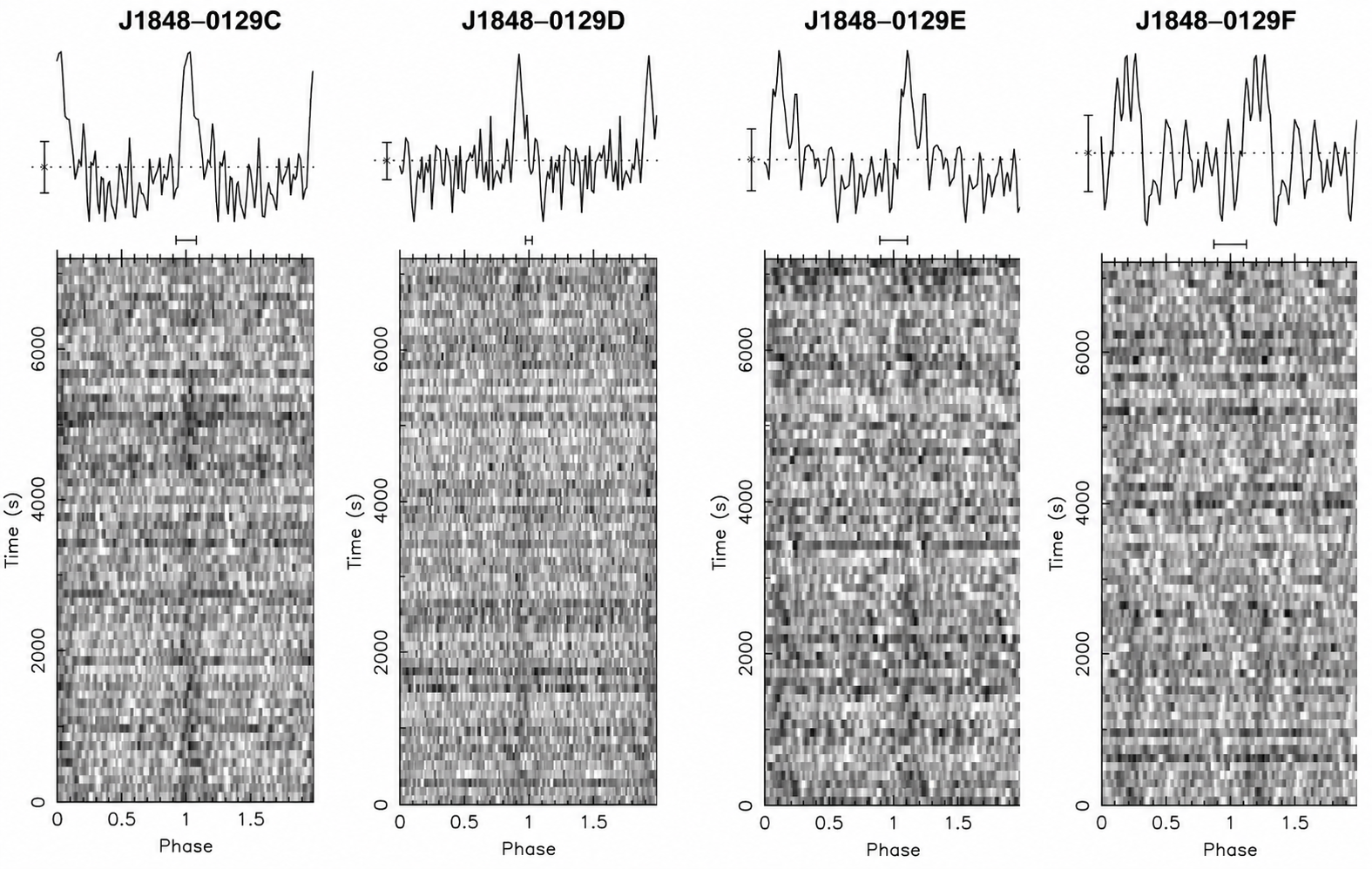}
    \caption{Time-versus-phase behaviour of the four newly discovered Glimpse-C01 pulsars. Each panel shows two full pulsar rotations for J1848$-$0129C, J1848$-$0129D, J1848$-$0129E, and J1848$-$0129F, shown in order.}
    \label{fig:phase_time}
\end{figure}

Four new pulsars were found in the GC Glimpse-C01 which had two previously known pulsars, J1848-0129A \citep{McCarver2024GlimpseC01} and J1848-0129B discovered by FAST \citep{2025ApJS..279...51L}. They are listed in Table~\ref{tab:new_pulsars}, together with the previously known pulsars. Time-versus-phase plots for these pulsars are shown in Fig.~\ref{fig:phase_time}.

This GC had already been searched in radio S-band (1850\,MHz) by the GBT and L-band by FAST. After the discovery of the four pulsars, the GBT archival data was searched again and the detection of J1848-0129D was confirmed. This was detected from acceleration searches performed on a 7.36 hour long observation. J1848-0129A was re-detected in the MeerKAT data but J1848-0129B, which was discovered by FAST could not be seen with blind searches on the original dataset (however, it was later detected using folding ephemerides provided by the FAST team).

After the discovery of the new pulsars in the MeerKAT data, Glimpse-C01 was observed on multiple occasions with MeerKAT in two dedicated orbital monitoring campaigns separated by approximately one year. These data, together with the GBT and FAST data, allowed some preliminary characterisation of these systems, which we now discuss.

\begin{table}[t]
\centering
\caption{Parameters of the pulsars in the GC Glimpse-C01.}
\small
\setlength{\tabcolsep}{3.5pt}
\renewcommand{\arraystretch}{0.92}
\label{tab:new_pulsars}
\begin{tabular}{l c c r}
\toprule
Pulsar & $P$ (ms) & DM ($\dmunit$) & $S_{2400}$ ($\mu$Jy) \\
\midrule
J1848$-$0129A & 19.78 & 491.1 & $32(2)$ \\
J1848$-$0129B & 13.80 & 482.0 & $10(1)$ \\
J1848$-$0129C & 6.44 & 491.8 & $35(2)$ \\
J1848$-$0129D & 17.10 & 457.9 & $12(1)$ \\
J1848$-$0129E & 4.54 & 479.9 & $8(1)$ \\
J1848$-$0129F & 4.17 & 520.1 & $8(1)$ \\
\bottomrule
\end{tabular}
\tablefoot{J1848$-$0129C, J1848$-$0129D, J1848$-$0129E and J1848$-$0129F are the four new discoveries reported in this work. The columns list the barycentric spin period $P$, dispersion measure DM, and mean MeerKAT S-band flux density $S_{2400}$. Uncertainties in parentheses correspond to $1\sigma$ errors on the last quoted digit(s); flux-density uncertainties are derived from the off-pulse noise of the folded profiles, as described in Sect.~\ref{sec:Flux_calc}.}
\end{table}

\subsubsection{J1848$-$0129C}
\label{sub:GlimpseC01-PsrC}

J1848$-$0129C is a 6.44\,ms binary MSP detected at a DM of $491.8\,\dmunit$. The pulsar was independently identified in both the 1\,km core and full-array PTUSE beams and was subsequently recovered in archival FAST observations.

The MeerKAT dataset spans a wide range of orbital phases and enabled the construction of a Keplerian orbital solution obtained by fitting a sinusoidal modulation to observed barycentric spin periods for J1848$-$0129C. The derived orbital parameters are given in Table~\ref{tab:j1848c_orbit}. The best-fitting model reproduces all available period measurements and the expected orbital modulation, as illustrated in Fig.~\ref{fig:phase_vs_spin}.

Assuming a neutron star mass of $\Mp = 1.4\,\msun$, the mass function implies a minimum companion mass of $M_{c,\min} = 0.3525\,\msun$ for an edge-on orbit ($i=90^\circ$) and a median companion mass of $\Mc = 0.4169\,\msun$ for the median inclination of $60^\circ$. The companion therefore lies in an intermediate mass regime, substantially heavier than the typical $\sim 0.23\,\msun$ helium white dwarf companions predicted for this orbital period by \cite{TS99}.

A notable property of J1848$-$0129C is its orbital-phase-dependent detectability. Using the solution in Table~\ref{tab:j1848c_orbit}, we computed the orbital phase corresponding to each MeerKAT observation. Detections are confined to phases away from superior conjunction, while observations predicted to sample phases near superior conjunction (orbital phase of 0.25 as seen in Fig.~\ref{fig:phase_vs_spin}) consistently result in non-detections.

In one case, two observations separated by less than 24\,h sampled orbital phases immediately preceding superior conjunction. The earlier observation yielded a detection, whereas the subsequent observation, probing deeper into the predicted conjunction phase interval, resulted in a non-detection. No clear ingress or egress structure is resolved within individual $\sim$2\,h integrations; however, the systematic confinement of non-detections to the phase range expected for superior conjunction is reproducible across campaigns separated by approximately one year.

The stability of this phase-locked non-detection interval disfavours refractive scintillation or acceleration-search limitations as the dominant cause of intermittency. Instead, the behaviour is consistent with eclipse or absorption in an intrabinary medium.

To assess whether the eclipse interpretation is geometrically plausible, we performed a simple consistency check using the derived orbital parameters. Adopting the phase interval $\Delta\phi\simeq0.24$ (phases $\sim0.13$--$0.37$) as the putative eclipse window, the corresponding duration is $\Delta t\simeq1.2$~d for $\Pb=5.03$~d. For a circular edge-on orbit, this phase width implies a minimum eclipsing scale $R_{\rm ecl}\gtrsim a\,\sin(\pi\Delta\phi)$.

Using Kepler’s third law with the parameters in Table~\ref{tab:j1848c_orbit} and $\Mp=1.4\,\msun$, the orbital separation is $a\simeq15\,\rsun$. Substituting $\Delta\phi\simeq0.24$ yields $R_{\rm ecl}\gtrsim10\,\rsun$ as a lower limit on the characteristic size of the absorbing region in the edge-on case. For comparison, the companion Roche-lobe radius for a mass ratio $q \equiv M_{\rm c}/M_{\rm p} \simeq 0.25$--0.30 is $R_{L,c}\simeq4\,\rsun$.

These order-of-magnitude estimates indicate that, if the non-detections are indeed caused by eclipse or absorption, the absorbing material must extend over several solar radii and cannot be explained by a compact geometric occultation alone. We emphasise that this calculation assumes a symmetric eclipse centred on superior conjunction and should be regarded as a preliminary geometric consistency check rather than a detailed physical model.

Classical redbacks are compact eclipsing MSP binaries with non-degenerate or semi-degenerate companions of mass $\sim 0.1$--$0.5\,M_\odot$ and orbital periods typically shorter than about a day \citep [see][]{Roberts2013}. In that interpretation, J1848$-$0129C is not compatible with such a compact redback configuration and instead occupies the long-period spider regime associated with the proposed `Huntsman' subclass \citep[see][]{2015ApJ...804L..12S, 2025ApJ...980..124S}. In particular, its orbital period places it directly alongside the Huntsman prototype PSR~J1417$-$4402 ($\Pb\simeq5.4$~d) and far beyond the orbital periods of known GC redbacks and candidates, whose longest currently reported value, held by J1740$-$5340B, is $\Pb\simeq1.97$~d \citep{2022ApJ...934L..21Z}.

If confirmed as a Huntsman-like eclipsing system, J1848$-$0129C would constitute the first example of this long-period spider subclass in a Galactic GC. Even under the more conservative classification of an `ordinary' redback, its $\Pb=5.03$~d would still represent the longest orbital period measured for a GC redback system by a wide margin.

\subsubsection{J1848$-$0129D}

J1848$-$0129D is a binary pulsar discovered in an acceleration search of MeerKAT S-band observations targeting the core of the GC Glimpse-C01 with the central 1 km array, and subsequently re-detected in archival GBT data as well as in FAST and later MeerKAT follow-up observations. The pulsar has a spin period of $P \simeq 17.1$~ms and a DM of $457.9\, \dmunit$.

\begin{figure*}[t]
    \centering
    \includegraphics[width=0.34\textwidth]{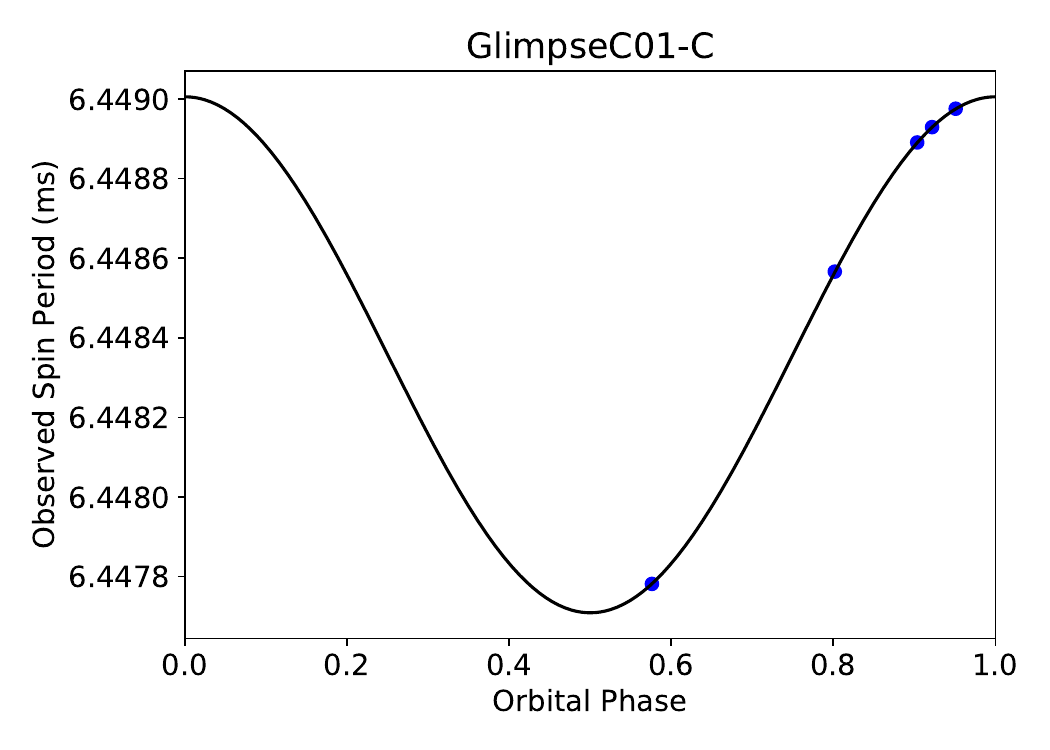}\hspace{0.035\textwidth}
    \includegraphics[width=0.34\textwidth]{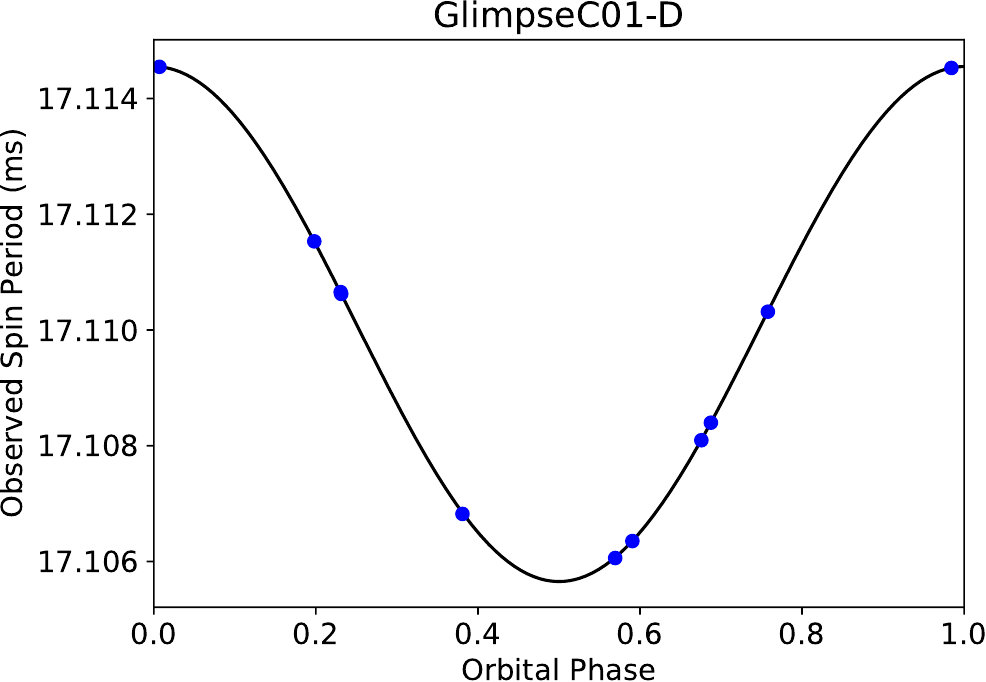}
    \caption{Observed spin-period modulation of J1848$-$0129C and J1848$-$0129D over orbital phase. Blue points show the observed spin periods, and solid black curves show the predictions calculated using the orbital solutions in Table~\ref{tab:j1848c_orbit}.}
    \label{fig:phase_vs_spin}
\end{figure*}

We obtained a preliminary Keplerian orbital fit by modelling the observed spin-period modulation using detections from MeerKAT. The resulting solution folds all available measurements, and the orbit is consistent with being circular ($e \approx 0$). The corresponding spin period modulation is also shown in Fig.~\ref{fig:phase_vs_spin}.

Assuming again $\Mp = 1.4 \, \msun$, the system's mass function implies $M_{c,\min}\simeq 0.98\,\msun$ and $\Mc\simeq 1.20\,\msun$ for the median inclination of $i=60^\circ$. While this mass range overlaps the lowest neutron-star masses, the absence of measurable eccentricity disfavours a double neutron star interpretation \citep[e.g.][]{2017ApJ...846..170T}; instead, the system is most naturally explained as a mildly recycled pulsar with a massive CO/ONeMg white dwarf companion formed through stable mass transfer \citep{2014MNRAS.437.1485L}.

\begin{table}[t]
\centering
\caption{Best-fitting orbital parameters for J1848$-$0129C and J1848$-$0129D.}
\label{tab:j1848c_orbit}
\scriptsize
\setlength{\tabcolsep}{2.5pt}
\renewcommand{\arraystretch}{0.91}
\begin{tabular}{l c c}
\toprule
Pulsar &  J1848$-$0129C & J1848$-$0129D \\
\midrule
Spin period, $P$ (ms) & 6.44835719(2) & 17.110102567(7) \\

Orbital period, $\Pb$ (d)                   & 5.0293642(6) & 3.39134255(7) \\
Projected semi-major axis, $x$ (lt-s)       & 6.95158(14) &  12.1271(4) \\
Epoch of periastron, $T_0$ (MJD) & 60591.731656(44) & 59279.22166(3) \\
Eccentricity, $e$ & 0 & 0 \\
Mass function, $f(M)$ ($\msun$) & $0.01426$ &  $0.16649$ \\
Minimum companion mass, $\Mc$ ($\msun$) & $0.3525$ & $0.9809$ \\
\bottomrule
\end{tabular}
\tablefoot{Uncertainties are quoted at the $1\sigma$ level and refer to the last quoted digit(s). The orbital parameters were obtained with a slightly modified version of \texttt{PRESTO}'s \texttt{fit\_circular\_orbit.py}.}
\end{table}

\subsubsection{J1848$-$0129E and J1848$-$0129F}

J1848$-$0129E and J1848$-$0129F are two faint binary MSPs. J1848$-$0129E was detected in the 1\,km core PTUSE beam and has a spin period of 4.54\,ms with a DM of $479.9\,\dmunit$, while J1848$-$0129F was detected in the full-array PTUSE beam and has a spin period of 4.17\,ms with a DM of $520.1\,\dmunit$. In the discovery observations, both pulsars were detected as accelerated candidates in the PRESTO acceleration searches, with best-fit line-of-sight accelerations of approximately \(a_{\rm los}\simeq 1.2~{\rm m\,s^{-2}}\) for J1848$-$0129E and \(a_{\rm los}\simeq 1.7~{\rm m\,s^{-2}}\) for J1848$-$0129F during the observation where they were discovered. These non-zero accelerations provide the present evidence that the two sources are candidate binary systems. However, the current detections are too limited to trace a coherent orbital modulation comparable to that shown for J1848$-$0129C and J1848$-$0129D.

Owing to their intrinsic faintness and the limited number of detections, no further constraints on the orbital parameters of either system could be obtained. No additional re-detections were secured during subsequent follow-up observations, preventing the derivation of coherent orbital solutions. At the time of writing, J1848$-$0129F has the highest dispersion measure reported for any pulsar associated with a GC. Further targeted observations will be required to confirm the binary properties of these systems and to better characterise their emission and propagation properties.

\subsection{Dispersion measure discrepancy and cluster membership}
\label{subsec:dm_discrepancy_membership}

Two of the pulsars detected towards Glimpse-C01 exhibit dispersion measures that differ substantially from the nominal value measured towards the cluster core ($\mathrm{DM}\simeq 491.1\,\dmunit$). In particular, J1848$-$0129D is detected at $\mathrm{DM}\simeq 457.9\,\dmunit$, while J1848$-$0129F is detected at $\mathrm{DM}\simeq 520.1\,\dmunit$, this means that the DMs of these pulsars have a full range of $61\,\dmunit$. This record DM range motivates a careful examination of whether the observed DM discrepancies are compatible with a cluster association, or whether we might be detecting MSPs unassociated with the cluster.

Several observational considerations support the interpretation that both pulsars are associated with Glimpse-C01 despite their displaced DMs. First, both detections were obtained in tied-array beams centred on the cluster core: J1848$-$0129D was detected in the 1-km core beam, while J1848$-$0129F was detected in the full-array beam. Their localisation within the cluster-centred beam coverage substantially reduces the probability of chance alignment with unrelated field pulsars.

If the pulsars are indeed associated with Glimpse-C01, the most natural explanation for the DM differences is real structure in the ionised interstellar medium along the line of sight. Glimpse-C01 lies behind a heavily obscured Galactic-plane sightline, where significant small-scale variations in electron column density are expected \citep[e.g.][]{CordesLazio2002,Yao2017}. In such environments, different sightlines across the angular extent of a globular cluster can sample distinct foreground ionised structures, leading to measurable DM differences among pulsars bound to the same cluster potential.

To place the Glimpse-C01 dispersion-measure behaviour in a broader context, we compared its DM distribution with those of other Galactic globular clusters hosting multiple known pulsars. Published dispersion measures were compiled from the MPIfR Globular Cluster Pulsar Catalogue, and only clusters with $\geq 5$ known pulsars were included to ensure that intracluster statistics are meaningful. For each cluster, we adopted the median pulsar DM as the nominal cluster DM and calculated $\sigma_{\rm DM}$ as the sample standard deviation of the individual pulsar DMs. We therefore used the $N-1$ denominator rather than $N$: division by $N$ would be appropriate if the measured pulsars represented the complete cluster pulsar population and its true mean DM were known, whereas the currently known pulsars constitute only a finite sample and their mean DM must be estimated from those same measurements. Using $N-1$ corrects for this lost degree of freedom and reduces the tendency to underestimate the underlying DM dispersion, particularly for clusters with relatively few measured pulsars. Only clusters with at least five known pulsars were included.

\begin{figure*}[t]
\sidecaption
\includegraphics[width=8.6cm]{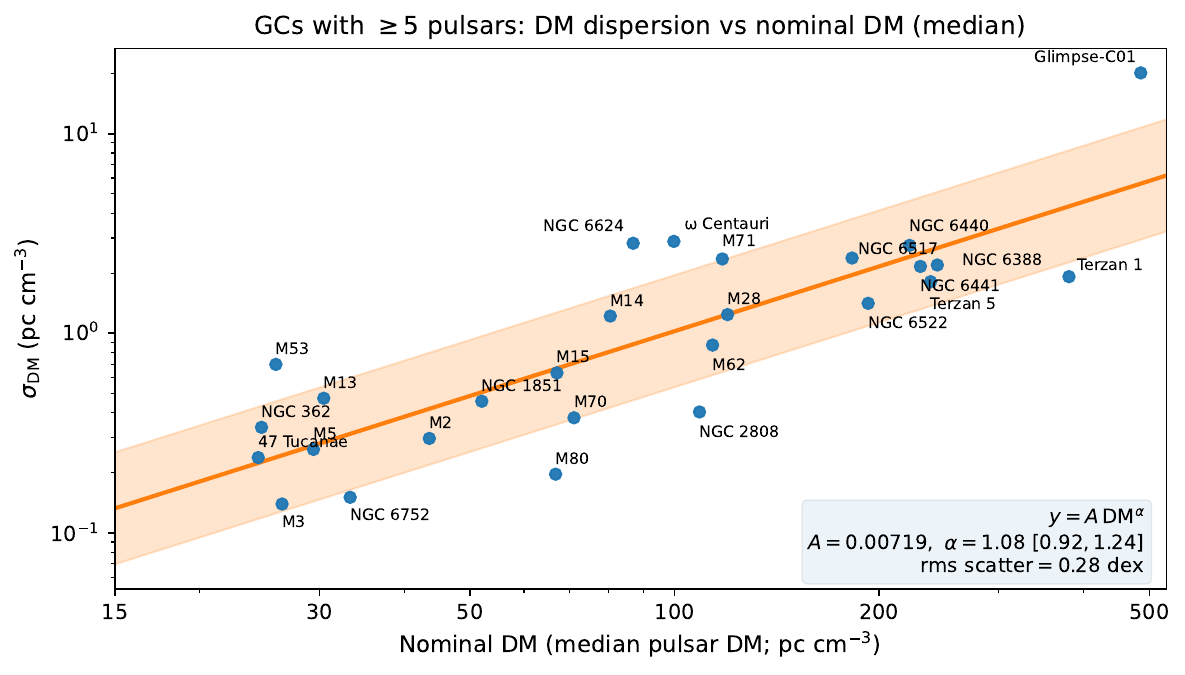}
\caption{Intracluster DM dispersion, $\sigma_{\rm DM}$, as a function of the nominal cluster DM, defined as the median pulsar DM, for Galactic GCs hosting at least five known radio pulsars. Blue points show the individual clusters. The solid orange line shows the best-fitting power-law relation $\sigma_{\rm DM}=A\,{\rm DM}^{\alpha}$, with $A=0.00719$ and $\alpha=1.08$ (bootstrap 16th--84th percentile interval: $0.92$--$1.24$). The orange shaded region represents the rms scatter of the residuals in log space, equal to $0.28$ dex. Pulsar DMs were compiled from the MPIfR Globular Cluster Pulsar Catalogue, accessed on 20 July 2026.}
\label{sigmaDM_vs_nominalDM_GCs}
\end{figure*}

Figure~\ref{sigmaDM_vs_nominalDM_GCs} shows $\sigma_{\rm DM}$ as a function of nominal cluster DM. A power-law relation of the form $\sigma_{\rm DM} = A\,\mathrm{DM}^{\alpha}$ was fitted in log--log space using ordinary least squares. To estimate uncertainties on the fitted parameters, we performed bootstrap resampling of the cluster sample (4000 realisations), refitting the relation for each resample and deriving 16th–84th percentile intervals for the slope and normalisation. The shaded region in the figure represents the rms scatter of the residuals in log space about the best-fitting relation.

A clear trend was observed: clusters with larger nominal DMs tend to exhibit larger intracluster DM dispersions. The best-fitting slope is statistically consistent with unity, indicating an approximately linear scaling $\sigma_{\rm DM} \propto \mathrm{DM}$. This behaviour is qualitatively consistent with the trend noted by \cite{Freire2005}, who investigated the spread of DMs in GCs using the full DM span, $\Delta\mathrm{DM} = \mathrm{DM}_{\max} - \mathrm{DM}_{\min}$, as a proxy for internal variation. At the time, the number of clusters hosting multiple known pulsars was limited, and the rms dispersion could not be robustly determined for most systems. With the substantially expanded cluster pulsar population now available, the use of $\sigma_{\rm DM}$ provides a more statistically stable measure of intracluster DM variation.

Glimpse-C01 lies at the high-DM end of the distribution and exhibits one of the largest internal DM dispersions in the sample. Using the six currently known pulsars in the cluster, we obtain
\(\sigma_{\rm DM}\simeq 20.3~{\rm pc~cm^{-3}}\) at a median DM of
\(\simeq 486.6~{\rm pc~cm^{-3}}\). The best-fitting relation shown
in Fig.~3 predicts \(\sigma_{\rm DM}\simeq 5.7~{\rm pc~cm^{-3}}\)
at this DM, so Glimpse-C01 lies about \(0.55\) dex above the best-fitting relation, corresponding to approximately twice the rms scatter of the residuals. It should therefore be regarded as an upper-envelope object within the high-DM cluster population, rather than as an isolated outlier requiring a different interpretation.

Taken together, the beam-centred detections and the placement of Glimpse-C01 within established population trends of intracluster DM dispersion support the interpretation that the DM offsets observed for J1848$-$0129D and J1848$-$0129F arise from small-scale structure in the intervening interstellar medium along this high-DM Galactic-plane sightline. By themselves, the observed offsets therefore do not require invoking a foreground or background origin.

\subsubsection*{Position of J1848$-$0129A}

The S-band survey observations of Glimpse-C01 included two dedicated tied-array beams placed on the reported position of J1848$-$0129A, one formed with the MeerKAT 1\,km core and the other with the full array. This reported position comes from the work of \citet{McCarver2024GlimpseC01}, where the VLA Low-band Ionosphere and Transient Experiment (VLITE) imaging on the Very Large Array identified a steep-spectrum point source as a plausible pulsar candidate, and a subsequent re-search of archival S-band data centred on the cluster led to the detection of J1848$-$0129A. The consistency of the measured flux densities led to the suggestion that the interferometric source and the pulsar were the same object. In our observations, however, J1848$-$0129A was not detected in either of the dedicated beams placed on that position, and it was instead detected in both the 1\,km core and full-array beams pointed at the cluster centre. This mismatch suggests that the reported interferometric position may not correspond to the true position of J1848$-$0129A, and that the steep-spectrum VLITE/VLA source could instead be unrelated to the pulsar.

This result was confirmed by a follow-up observation of Glimpse-C01 conducted in April 2024 as part of an orbital determination and timing campaign for the newly discovered cluster pulsars. These observations used the same beam configurations as the original S-band survey. Once again, J1848$-$0129A was not detected in the two beams centred on the VLITE position, but it was clearly detected in the beams pointed at the cluster core, reinforcing the conclusion that the VLITE point source is unlikely to be associated with J1848$-$0129A.

To further investigate if the VLITE point source is truly a new unidentified pulsar, we performed full as well as segmented acceleration and jerk searches on both the original S-band survey data and the April 2024 follow-up observations at the VLITE beam pointings. These searches were carried out using both the 1\,km core and full-array data, using very high acceleration and jerk trial values. No additional pulsars were detected in any of these searches. As a result, the radio source detected in the VLITE survey\footnote{\url{https://vlite.nrao.edu/}}, originally associated with J1848$-$0129A, remains unidentified.

\subsection{S-band and L-band flux density measurements}
\label{Flex_density_measurements}

Table~\ref{tab:flux_simple} presents mean flux density measurements for a subset of pulsars detected in the MeerKAT S-band GC survey that also have available MeerKAT L-band observations. The pulsars listed were selected based on the availability of reliable ephemerides, which allowed coherent folding of both the S-band survey data and archival or targeted L-band observations. The S-band flux densities reported here were derived from single-epoch observations, whereas the corresponding L-band flux densities were obtained by averaging measurements from multiple epochs where such data were available.

For pulsars with flux-density measurements available in both bands, we estimated a two-point spectral index by assuming a single power-law spectrum, $S_{\nu}\propto\nu^{\alpha}$, between the representative L-band and S-band frequencies. These values should therefore be regarded as two-frequency spectral-index estimates rather than broadband spectral fits, and they do not constrain possible spectral curvature or turnovers.

It is important to note that the S-band and L-band observations used to populate Table~\ref{tab:flux_simple} were not obtained contemporaneously. In several cases, the observations are separated by months to years. Over such timescales, refractive interstellar scintillation is expected to introduce substantial variability in the observed flux densities \citep[see e.g.][]{2022ApJ...941...22M}, potentially enhancing or suppressing the measured signal by factors of a few \citep{Rickett1990,LorimerKramer2005}. The uncertainties quoted for the single-epoch MeerKAT S-band flux densities are not inflated to account for this epoch-to-epoch variability; they represent the formal measurement uncertainties for the observed epoch, derived from the folded-profile noise. By contrast, the quoted L-band values are multi-epoch averages where available, or literature values, and their uncertainties reflect the corresponding averaging or literature measurement procedure. As a result, the flux density values reported in the two bands are not intended to be compared directly, and any apparent differences between the L-band and S-band measurements should not necessarily be interpreted as indicative of intrinsic spectral behaviour.

The primary purpose of Table~\ref{tab:flux_simple} is therefore to provide reference flux density measurements for pulsars detected in the S-band survey. The use of multi-epoch averaging at L-band serves to mitigate the effects of scintillation and provides a more representative estimate of the typical L-band flux density for comparison at a population level. These measurements also act as a consistency check for the flux density estimation procedure described in Appendix~\ref{flux_density_method}, applied across different observing bands and datasets.

Terzan~5 is presented separately here because it contributes a particularly large number of pulsar detections within a single cluster, and because extensive literature flux density measurements are available for its known pulsars. Using the $S_{1400}$ and $S_{2000}$ values reported by \citet{2022ApJ...941...22M}, we compare these with our single-epoch MeerKAT S-band flux density measurements at $\sim 2406$\,MHz. Table~\ref{tab:ter5_flux_density} lists the resulting $S_{2400}$ values together with the literature measurements, and gives updated spectral indices derived using the literature $S_{1400}$ values and our measured $S_{2400}$ flux densities. This allows the Terzan~5 detections to be placed in the context of previous flux density measurements while keeping this unusually large single-cluster sample separate from the rest of the survey detections.

A more quantitative investigation of spectral properties would require either simultaneous multi-frequency observations or a sufficiently large number of epochs at each frequency to average over scintillation effects. Such an analysis is beyond the scope of the present work.

Beyond serving as reference measurements, the flux densities listed in Table~\ref{tab:flux_simple} provide an empirical indication of the range of pulsar brightnesses accessible to MeerKAT S-band observations in globular clusters. The detection of several known pulsars with reliable orbital solutions at 2.4\,GHz demonstrates the viability of S-band searches for both isolated and binary millisecond pulsars in dense cluster environments. These measurements therefore establish a practical sensitivity scale for the present survey, which is useful for interpreting subsequent non-detections and for informing the design of future high-frequency globular-cluster pulsar surveys.

\subsection{S-band non-detections and reference spectral-index constraints}
\label{sec:sband_nondetections}

A number of previously known globular-cluster pulsars fall within the MeerKAT S-band ($\sim$2.4\,GHz) PTUSE tied-array beam coverage but are not detected in our S-band survey. Several of these pulsars are, however, reliably detected in MeerKAT L-band ($\sim$1.4\,GHz) observations reported in the literature. While our S-band survey consists of a single epoch per cluster, we examine whether the combination of published L-band detections and S-band non-detections can be used to derive reference upper limits on the radio spectral index.

This exercise is not intended to yield precise spectral measurements. Instead, the derived limits are meant to serve as indicative constraints that help contextualise S-band non-detections of otherwise established L-band pulsars. The resulting values are expected to carry substantial uncertainties due to the limited number of epochs, potential epoch-to-epoch variability driven by interstellar scintillation, and assumptions regarding the effective sensitivity of coherent tied-array beams. For the purpose of deriving upper limits, we adopt the nominal usable S1-band bandwidth of 875\,MHz for the best case scenario. This is very close to the typical effective bandwidth after RFI excision, which is $\gtrsim 98\%$ of the processed band. The resulting limits, therefore, represent optimistic, best-case sensitivity estimates.

Of the full set of S-band non-detections, we restricted our attention to pulsars that (i) are reported to be of moderate to high brightness at L-band in published MeerKAT or comparable observations, (ii) are not known spider systems (black widows or redbacks) or strongly eclipsing binaries, and (iii) have sufficiently precise positional localisations to confirm that they lie within the S-band PTUSE beam response. Pulsars clearly outside the S-band beam footprint, as well as systems with poor L-band localisation for which beam coverage cannot be reliably assessed, were excluded from this analysis. This selection is intended to minimise ambiguity arising from eclipses, orbital modulation, and positional uncertainties.

Assuming a power-law radio spectrum,
\begin{equation}
    S_{\nu} \propto \nu^{\alpha},
\end{equation}
an S-band non-detection can be combined with a published L-band flux density to place an upper limit on the spectral index $\alpha$. The minimum detectable flux density at the S-band was estimated using the standard radiometer equation \citep{LorimerKramer2005},
\begin{equation}
S_{\rm min} =
{\rm S/N}_{\rm min}
\frac{\mathrm{SEFD}_{\rm array}}
{\sqrt{n_{\rm pol}\,\Delta\nu\,t_{\rm obs}}}
\sqrt{\frac{\delta}{1-\delta}},
\end{equation}
where ${\rm S/N}_{\rm min}$ is the adopted detection threshold, $\mathrm{SEFD}_{\mathrm{array}}$ is the effective system-equivalent flux density of the coherently summed tied-array beam\footnote{\url{https://skaafrica.atlassian.net/wiki/spaces/ESDKB/pages/1588854789/S-band+capability+and+status}}, $n_{\mathrm{pol}}=2$ is the number of summed polarisations, $\Delta\nu$ is the usable bandwidth, $t_{\mathrm{obs}}$ is the integration time, and $\delta=W/P$ is the pulse duty cycle. For consistency, we adopt duty cycles comparable to those inferred at L-band, noting that this is a simplifying assumption.

Given an L-band flux density $S_{\rm L}$ at frequency $\nu_{\rm L}$ and an S-band non-detection corresponding to an upper limit $S_{\rm min}$ at frequency $\nu_{\rm S}$, the associated upper limit on the spectral index is
\begin{equation}
\alpha_{\rm max} =
\frac{\ln\left(S_{\rm min}/S_{\rm L}\right)}
{\ln\left(\nu_{\rm S}/\nu_{\rm L}\right)}.
\end{equation}
In the following, we adopt $\nu_{\mathrm{S}}=2.4$\,GHz, $t_{\mathrm{obs}}=2$\,hr, $\Delta\nu=875$\,MHz, and ${\rm S/N}_{\rm min}=8$ for our S-band search configuration. The effective $\mathrm{SEFD}_{\mathrm{array}}$ is computed from the published MeerKAT S-band receiver-performance curves by dividing the band-averaged per-antenna SEFD by the number of antennas contributing to the coherent tied-array sum, as described in Appendix~\ref{app:sefd}.

Under these assumptions, the effective S-band detection threshold implied by the radiometer equation corresponds to a practical reference level of the order of 10--20~$\mu$Jy at 2.4\,GHz for the 1\,km PTUSE beams. For non-detections of pulsars with known positions, the effective sensitivity may be reduced relative to the ideal radiometer-equation prediction if the pulsar lies away from the beam boresight. Rather than modelling the detailed tied-array beam response and epoch-specific conditions for each case, we adopt a conservative sensitivity margin when deriving spectral-index limits. This approach is intended to bracket plausible detectability while avoiding over-interpretation of single-epoch non-detections.

Among the non-detections satisfying the selection criteria described above, only J1823$-$3021G and J1824$-$2452D have sufficiently reliable L-band flux density measurements and positional information to permit meaningful spectral-index constraints. We therefore restricted the following analysis to these two systems.

For J1823$-$3021G, \citet{Ridolfi2021} report a mean L-band flux density of $S_{1300} = 47\pm10~\mu\mathrm{Jy}$. Combining this with the S-band non-detection yields an indicative constraint of $\alpha \lesssim -1.6$ to $-1.2$, where the range reflects conservative sensitivity assumptions and the quoted uncertainty on the L-band flux density.

A similar constraint can be derived for J1824$-$2452D, which lies within our 1\,km S-band PTUSE beam coverage but is not detected in our S-band data. A published L-band reference flux density for J1824$-$2452D is $S_{1400}\simeq50\,\mu{\rm Jy}$ \citep{2011MNRAS.418..477B}. Under the same conservative assumptions, the S-band non-detection implies an upper limit on the spectral index in the range $\alpha \lesssim -2.5 $ \text{to} $ -2.1$.

A similar consideration applies to several of the non-detections in Terzan~5. Using the literature flux densities and spectral indices  \citep{2022ApJ...941...22M}, extrapolated to our observing frequency of $\sim 2406$\,MHz, we find that PSRs~J1748$-$2446H, J1748$-$2446R, J1748$-$2446U, J1748$-$2446ac, J1748$-$2446aa, J1748$-$2446ag, and J1748$-$2446ah are all expected to be very faint at MeerKAT S band, and their non-detection in our single-epoch observations is therefore not surprising. In the same context, PSRs~J1748$-$2446ao and J1748$-$2446aw are already very faint in MeerKAT L-band observations  \citep{2024A&A...686A.166P}, with $S_{1284}=12$\,$\mu$Jy and $10$\,$\mu$Jy, respectively, and their absence in our S-band data is likewise expected, indicating no evidence of an unusual inverted spectrum in either case. By contrast, the extrapolated S-band flux densities of PSRs~J1748$-$2446F and J1748$-$2446K are $\sim29$\,$\mu$Jy and $\sim30$\,$\mu$Jy, respectively, which places them within the practical sensitivity range of our MeerKAT S-band observations. Since both are isolated pulsars, their non-detection is unlikely to be related to orbital-phase effects, and is more plausibly explained by refractive scintillation.

Taken at face value, these limits imply that the pulsars would need to exhibit relatively steep radio spectra for their published L-band flux densities to fall below the S-band detection threshold at 2.4\,GHz. In practice, single-epoch non-detections can also be influenced by refractive scintillation and possible duty-cycle evolution with observing frequency. The derived values should therefore be interpreted as conservative upper bounds intended to characterise expected detectability in a single-epoch S-band survey.

\begin{table}[t]
\centering
\caption{MeerKAT L- and S-band flux densities and two-point spectral indices.}
\scriptsize
\setlength{\tabcolsep}{3.2pt}
\renewcommand{\arraystretch}{0.88}
\label{tab:flux_simple}
\begin{tabular}{llccc}
\toprule
GC & Pulsar name & $S_{1200}$ (mJy) & $S_{2400}$ (mJy) & $\alpha$ \\
\midrule
M22 & J1836$-$2354A & 0.201(23) & 0.074(12) & $-1.44(29)$ \\
    & J1836$-$2354B & 0.128(45) & 0.071(11) & $-0.85(55)$ \\

NGC6522 & J1803$-$3002A & 0.318(51) & 0.094(15) & $-1.76(33)$ \\

NGC6539 & J1804$-$0735A & -- & 0.090(14) & -- \\

M28 & J1824$-$2452A & 0.328(53) & 0.192(31) & $-0.77(33)$ \\
    & J1824$-$2452B & 0.019(3)  & 0.012(2)  & $-0.66(33)$ \\
    & J1824$-$2452C & 0.089(14) & 0.078(12) & $-0.19(32)$ \\
    & J1824$-$2452E & --        & 0.014(2)  & -- \\
    & J1824$-$2452G & --        & 0.032(5)  & -- \\

NGC6712 & J1853$-$0842A & -- & 0.024(4) & -- \\

NGC6624 & J1823$-$3021A & 0.218(340) & 0.124(34) & $-0.81(228)$ \\
        & J1823$-$3021B & 0.092(21)  & 0.084(14) & $-0.13(41)$ \\
        & J1823$-$3021D & 0.051(19)  & 0.040(14) & $-0.35(74)$ \\
\bottomrule
\end{tabular}
\tablefoot{The $S_{1200}$ values are averages over multiple L-band epochs, subject to data availability, whereas the $S_{2400}$ values are derived from single-epoch S-band detections. Spectral indices are computed assuming $S_{\nu}\propto\nu^{\alpha}$ between 1.2 and 2.4\,GHz. Uncertainties are quoted at the $1\sigma$ level; values in parentheses denote the uncertainty on the last quoted digit(s).}
\end{table}

\begin{table}[t]
\centering
\caption{Flux-density measurements for the Terzan~5 pulsars detected at S-band.}
\scriptsize
\setlength{\tabcolsep}{3.2pt}
\renewcommand{\arraystretch}{0.88}
\label{tab:ter5_flux_density}
\begin{tabular}{lccccc}
\hline
\hline
Pulsar name & $S_{1400}$ (mJy) & $S_{2000}$ (mJy) & $S_{2400}$ (mJy) & $\alpha$ \\
\hline
J1748$-$2446A  & 2.700 & 1.700 & 0.702(9)  & $-2.50$ \\
J1748$-$2446C  & 1.100 & 0.670 & 0.613(12) & $-1.08$ \\
J1748$-$2446E  & 0.170 & 0.110 & 0.090(10) & $-1.18$ \\
J1748$-$2446G  & 0.024 & 0.022 & 0.027(2)  & $+0.22$ \\
J1748$-$2446I  & 0.095 & 0.055 & 0.034(1)  & $-1.91$ \\
J1748$-$2446L  & 0.096 & 0.043 & 0.026(9)  & $-2.42$ \\
J1748$-$2446M  & 0.140 & 0.091 & 0.068(16) & $-1.34$ \\
J1748$-$2446N  & 0.150 & 0.100 & 0.097(22) & $-0.81$ \\
J1748$-$2446O  & 0.310 & 0.160 & 0.101(30) & $-2.08$ \\
J1748$-$2446V  & 0.100 & 0.077 & 0.034(11) & $-2.00$ \\
J1748$-$2446W  & 0.054 & 0.031 & 0.022(8)  & $-1.67$ \\
J1748$-$2446Y  & 0.037 & 0.029 & 0.021(7)  & $-1.05$ \\
J1748$-$2446Z  & 0.030 & 0.023 & 0.019(7)  & $-0.85$ \\
J1748$-$2446ae & 0.056 & 0.050 & 0.043(11) & $-0.49$ \\
J1748$-$2446ai & 0.033 & 0.028 & 0.021(11) & $-0.84$ \\
J1748$-$2446am & \dots & \dots & 0.008(2)  & \dots \\
\hline
\end{tabular}
\tablefoot{Owing to the large number of pulsars detected in Terzan~5, these measurements are presented separately from the rest of the sample. The $S_{1400}$ and $S_{2000}$ values are taken from \citet{2022ApJ...941...22M}, while $S_{2400}$ gives our measured flux densities at a central frequency of $\sim2406$\,MHz. The spectral index $\alpha$ is computed from the literature $S_{1400}$ value and our measured $S_{2400}$ assuming $S_{\nu}\propto\nu^{\alpha}$. No literature flux-density measurement was available for PSR~J1748$-$2446am.}
\end{table}

\section{Discussion and conclusions}
\label{sec:disc_conc}

\subsection{S-band sensitivity and implications for future surveys}

With this work, we have presented the first dedicated MeerKAT S-band ($\nu \simeq 2.4$\,GHz) pulsar survey of GCs. The survey targeted 14 systems observed during 2023--2024. The survey re-detected 39 previously known pulsars and discovered four new millisecond pulsars in Glimpse-C01 (J1848$-$0129C, D, E, and F), demonstrating that coherent tied-array beams at the S-band provide a powerful complement to lower-frequency searches, particularly for high-DM clusters where dispersion and scattering increasingly limit detectability at $\sim$1--1.4\,GHz.

A key practical outcome of the survey is an empirical single-epoch detectability scale for MeerKAT S-band tied-array observations. Across detections and non-detections, the results are consistent with a characteristic sensitivity of the order of 10--20~$\mu$Jy for $\sim$2\,hr integrations under the survey configuration adopted here. This level is best interpreted as a practical single-epoch sensitivity reference for comparable MeerKAT S-band tied-array observations with similar dwell times and instrumental setup, rather than as a strict completeness threshold. The associated search strategy is more specific to targeted observations such as GCs, where prior knowledge of the approximate DM allows the residual DM range to be restricted and more computational effort to be devoted to acceleration, jerk, and segmented searches.

The results are particularly relevant to the ongoing MeerKAT+ extension, which will augment MeerKAT with more than a dozen SKA-Mid-design dishes equipped with both Band-2 and MPIfR S-band receivers. The latter retain the 1.75--3.5 GHz frequency coverage used in this survey, while the enlarged array will provide increased collecting area and hence improved sensitivity to faint pulsars. MeerKAT+ therefore offers a direct route towards deeper S-band searches of highly dispersed and scattered GCs, particularly for short integrations designed to retain sensitivity to compact binary systems. The new dishes are ultimately intended to be incorporated into SKA-Mid, which will allow the observing and processing strategies demonstrated here to inform future high-frequency pulsar-search capabilities.

\subsection{The newly discovered Glimpse-C01 pulsars}

J1848$-$0129C exhibits a combination of a multi-day orbit and reproducible orbital-phase-dependent detectability. The orbital solution yields $\Pb=5.03$~d and a companion mass of $\Mc\simeq0.35$--$0.42\,\msun$ for typical inclinations. Non-detections are confined to orbital phases near superior conjunction and recur across campaigns separated by approximately one year. This phase-locked behaviour disfavours stochastic propagation effects and is suggestive of eclipse or absorption within the binary system. Interpreting the phase interval $\Delta\phi\simeq0.24$ as an eclipse window corresponds to an eclipse duration of $\sim1.2$~d. Under the assumption of a circular edge-on orbit, this phase width implies a characteristic absorbing scale of the order of $\sim10\,\rsun$, based on an orbital separation of $a\simeq15\,\rsun$. For comparison, the expected Roche-lobe radius of a $\sim0.4\,\msun$ companion in such an orbit is $\sim4\,\rsun$. While these estimates are intended as approximate geometric constraints, they indicate that a purely compact stellar occultation would be insufficient to account for an eclipse of this duration, and that extended intrabinary material would likely be required if the eclipse interpretation is correct.

With $\Pb=5.03$~d, J1848$-$0129C lies at the extreme long-period end of the spider population, typically showing $<1$~d orbits. Its orbital period exceeds that of known GC redbacks and is comparable to that of the proposed `Huntsman' systems, such as PSR~J1417$-$4402. An alternative possibility is that the present companion is a main-sequence dwarf star, as expected for ordinary cluster stars with masses $\lesssim 0.7\,\msun$, that replaced the original recycling companion in a later exchange interaction. Such a scenario could naturally produce Huntsman-like properties, with the MSP now interacting with a non-degenerate companion, while tidal effects may also have contributed to circularising the present orbit. Further multi-wavelength observations will be necessary to determine whether the companion is non-degenerate or evolved, and to establish whether J1848$-$0129C represents a long-period spider in a GC or an unusually wide detached configuration.

The orbital parameters of J1848$-$0129D have also been obtained. Using detections from MeerKAT, we obtained a preliminary Keplerian orbital fit by modelling the observed spin-period modulation. The resulting solution correctly folds all available data, and the orbit is consistent with being circular. The resulting companion mass, assuming $\Mp =1.4\,\msun$, lies in the range $M_{\rm c} \simeq 0.98$--$1.20\,\msun$ for inclinations $i$ between $90^{\circ}$ and $60^{\circ}$. While this mass range overlaps the lowest measured neutron-star masses, the absence of measurable eccentricity disfavours a double-neutron-star interpretation and instead points to a mildly recycled pulsar with a massive CO/ONeMg white dwarf companion formed via stable mass transfer. Continued phase-connected timing will refine the orbital parameters and could enable detection of post-Keplerian effects.

J1848$-$0129E and J1848$-$0129F are fainter binary MSPs for which the current detection set is insufficient to obtain coherent orbital solutions. All of the Glimpse-C01 pulsars are also being pursued in ongoing timing efforts. A timing analysis of J1848$-$0129A is in preparation (McCarver et al.\ in prep.), as is timing work on J1848$-$0129B (Baoda et al.\ in prep.). Timing work on the newly discovered systems, including J1848$-$0129C and J1848$-$0129D, is also in preparation (Nag et al.\ in prep.).

One notable feature of the pulsars discovered to date in Glimpse-C01 is that half of them have spin periods between 10 and 20 ms, i.e. they are relatively mildly recycled. In the case of J1848$-$0129D, this can be understood from the fact that it has a massive WD companion, as its spin period agrees with the observed spin periods of pulsars with similarly massive WD companions in the Galactic disk. We note that the low eccentricity does not hint at exchange interactions in the history of this system. This raises the question of whether slow pulsars with massive companions are over-abundant in this GC. If so, then something about the evolution of this GC is unusual. However, it is likely that dispersive smearing and multipath scattering in this GC may be preventing the detection of many of the faster-spinning pulsars. If this is the case, then the pulsar population of Glimpse-C01 might be more similar to that of other GCs with a similar encounter rate, which would imply a large number of faster pulsars still to be discovered.

\subsection{Dispersion-measure structure and survey strategy}

GLIMPSE-C01 is one of the most highly obscured GCs in the Milky Way, with a visual extinction of $A_V \sim 18$ magnitudes \citep{Hare2018}, indicating that the cluster lies behind a very large column of intervening material. It is reported to exhibit pronounced spatial variations in optical extinction across its extent (strong differential reddening; Cadelano et al., in prep.). Among clusters hosting identified pulsars, it is the most heavily extinct system currently known. The combination of a high mean DM and a large internal DM dispersion may therefore be consistent with substantial small-scale structure in the intervening material along neighbouring sightlines. This raises the possibility of a connection between fluctuations in ionised gas traced by DM and the complex dust distribution probed at optical and near-infrared wavelengths (Pallanca et al., in prep.), although a quantitative assessment of such a link lies beyond the scope of the present work.

The DM behaviour observed towards Glimpse-C01 has direct implications for survey strategy at high foreground DMs. Two pulsars detected towards the cluster, J1848$-$0129D and J1848$-$0129F, exhibit substantial offsets from the median cluster DM. A population comparison across GCs with $\geq 5$ known pulsars shows that Glimpse-C01 lies at the high-DM extreme and follows the empirical trend in which larger median DMs are associated with larger intracluster DM spreads (see Fig. \ref{sigmaDM_vs_nominalDM_GCs}). Practically, this motivates adopting broader (or adaptive) DM search padding for high-DM clusters, since narrowly padded DM windows become increasingly likely to miss genuine cluster pulsars in strongly structured lines of sight. Finally, the S-band and L-band flux-density measurements reported here provide a uniform high-frequency reference set for re-detected pulsars and a basis for illustrative spectral-index constraints in selected non-detections. The combined results show that MeerKAT S-band tied-array observations are capable of recovering faint MSPs in dense cluster environments and open a complementary discovery space for highly dispersed systems that are challenging at lower frequencies.

\section*{Data availability}

The data underlying this work are held by the author and are stored on the HPC facilities of the INAF--Osservatorio Astronomico di Cagliari. They will be made available upon reasonable request.

\begin{acknowledgements}
We thank the staff of the South African Radio Astronomy Observatory (SARAO) for their support of MeerKAT operations. The MeerKAT telescope is operated by SARAO, which is a facility of the National Research Foundation, an agency of the Department of Science, Technology and Innovation. This work has made use of the `MPIfR S-band receiver system' designed, constructed and maintained by funding of the Max-Planck-Institut f\"ur Radioastronomie and the Max-Planck-Gesellschaft. Observations made use of the Pulsar Timing User Supplied Equipment (PTUSE) servers at MeerKAT which were funded by the MeerTime Collaboration members ASTRON, AUT, CSIRO, ICRAR--Curtin, MPIfR, INAF, NRAO, Swinburne University of Technology, the University of Oxford, UBC and the University of Manchester.

The MeerKAT S-band census data products used in this work were hosted on the OzSTAR national facility at Swinburne University of Technology, where data subbanding and format conversion to \texttt{.fits} were performed. The OzSTAR programme receives funding in part from the Astronomy National Collaborative Research Infrastructure Strategy (NCRIS) allocation provided by the Australian Government.

The National Radio Astronomy Observatory is a facility of the National Science Foundation operated under cooperative agreement by Associated Universities, Inc. Archival confirmations of J1848$-$0129D made use of Green Bank Telescope data. The Green Bank Observatory is a facility of the National Science Foundation operated under cooperative agreement by Associated Universities, Inc. We also thank Z.~Pan and collaborators for assistance with archival FAST data that helped confirm the detections of J1848$-$0129C and J1848$-$0129D; FAST is a Chinese national mega-science facility operated by the National Astronomical Observatories of the Chinese Academy of Sciences.

We acknowledge the MeerTime and TRAPUM collaborations for enabling access to observations and data products used in this work. We thank the University of Cagliari and the INAF--Osservatorio Astronomico di Cagliari for institutional support and for providing HPC and long-term storage resources that enabled the analysis.

The INAF--OAC computer cluster used in this work has been acquired within a project aimed at enhancing the Sardinia Radio Telescope (SRT). The enhancement of the SRT for the study of the Universe at high radio frequencies is financially supported by the National Operative Program (Programma Operativo Nazionale -- PON) of the Italian Ministry of University and Research `Research and Innovation 2014--2020', Notice D.D. 424 of 28/02/2018 for the granting of funding aimed at strengthening research infrastructures, in implementation of Action II.1 -- Project Proposal PIR01\_00010.

We are grateful to M.~Bailes for sharing code and guidance used in implementing the on-pulse flux-density estimation approach adopted in this paper. This work was supported in part by the Italian Ministry of Foreign Affairs and International Cooperation, grant number ZA23GR03, under the project `RADIOMAP -- Science and technology pathways to MeerKAT+: the Italian and South African synergy'.
A.P., M.B., and M.C.iB. acknowledge the help obtained from the resources provided by the INAF Large Grant 2022 `GCjewels' (P.I. Andrea Possenti) approved with the Presidential Decree 30/2022. S.M.R.~is a CIFAR Fellow and is supported by the NSF Physics Frontiers Center award 2020265.

The work of the leading author was supported by the Fondazione ICSC, Spoke~3 Astrophysics and Cosmos Observations.
National Recovery and Resilience Plan (Piano Nazionale di Ripresa e Resilienza, PNRR) Project ID CN\_00000013 `Italian Research Center on High-Performance Computing, Big Data and Quantum Computing' funded by MUR Missione~4 Componente~2 Investimento~1.4: Potenziamento strutture di ricerca e creazione di `campioni nazionali di R\&S (M4C2-19)' -- Next Generation EU (NGEU).

The lead author acknowledges that the research activities described in this paper were carried out with contribution of the NextGenerationEU funds within the National Recovery and Resilience Plan (PNRR), Mission~4 -- Education and Research, Component~2 -- From Research to Business (M4C2), Investment Line~3.1 -- Strengthening and creation of Research Infrastructures, Project IR0000034 -- `STILES -- Strengthening the Italian Leadership in ELT and SKA'. VVK acknowledges continuing support from the Max Planck Society and financial support from the European Research Council (ERC) starting grant `COMPACT' (Grant agreement number 101078094).
\end{acknowledgements}

\bibliographystyle{aa}
\bibliography{references}

\begin{appendix}
\begin{figure*}[!t]
\section{Survey overview material}
\label{appendix}
\centering
\includegraphics[width=0.195\textwidth]{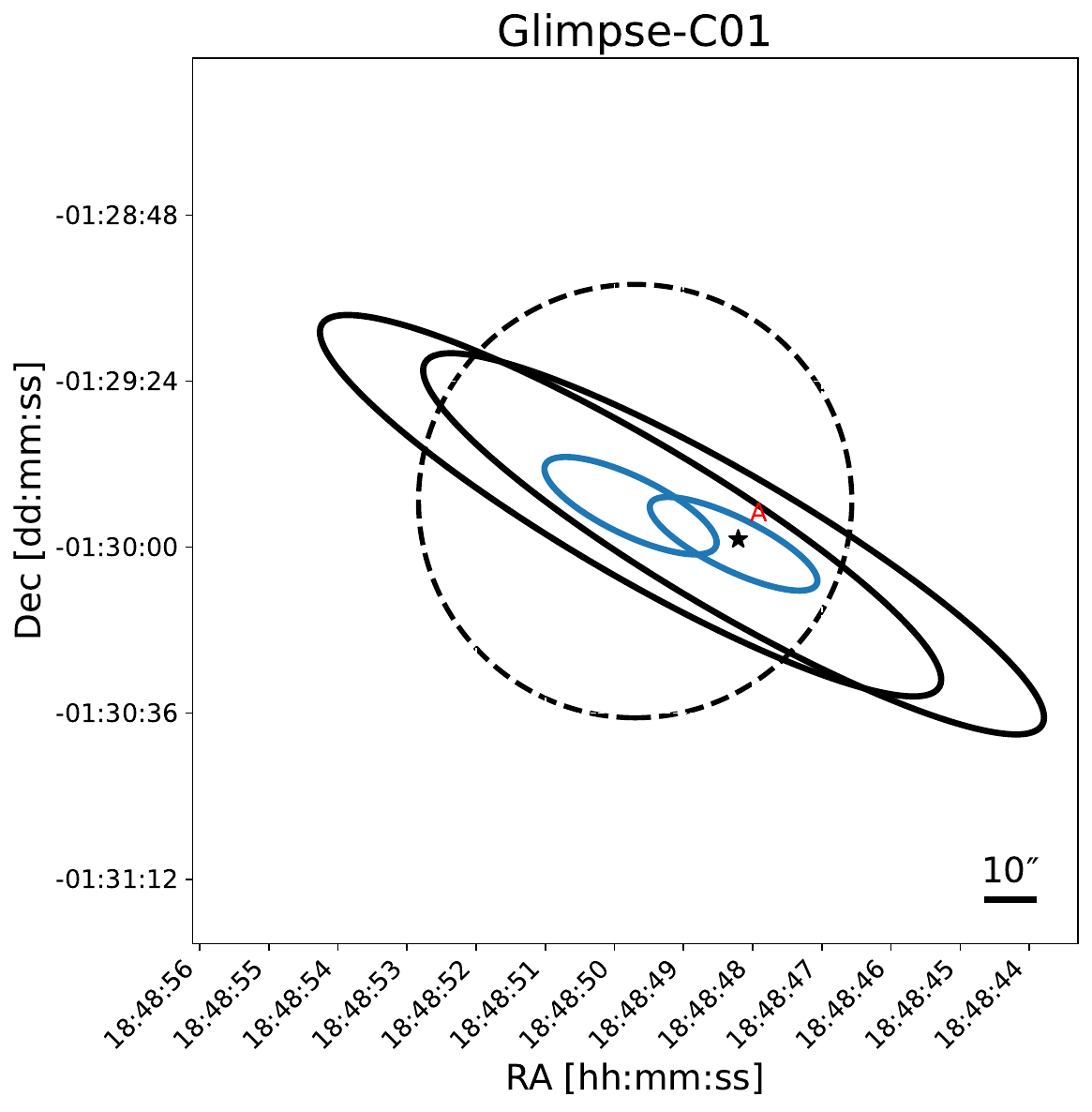}\hfill
\includegraphics[width=0.195\textwidth]{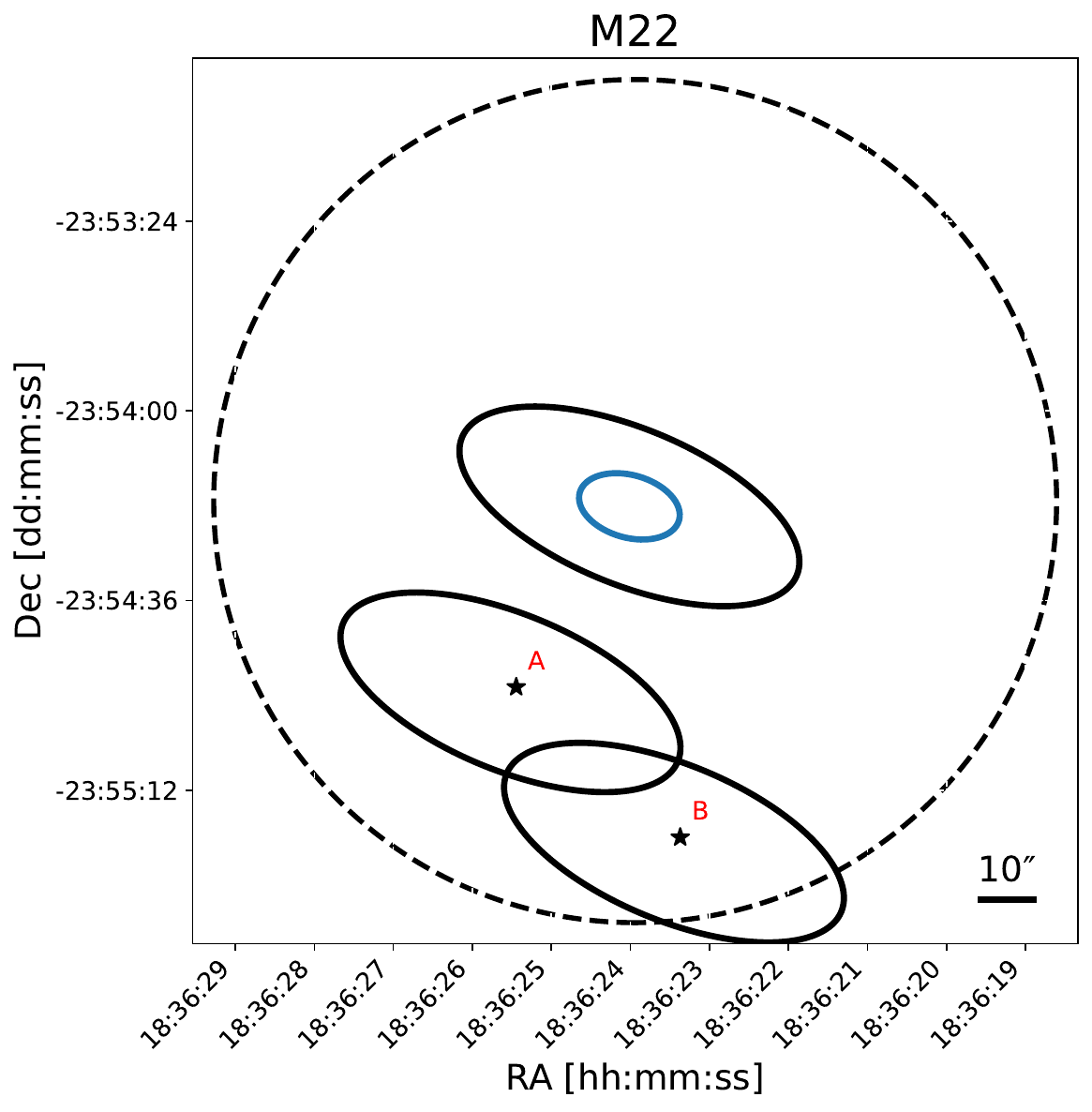}\hfill
\includegraphics[width=0.195\textwidth]{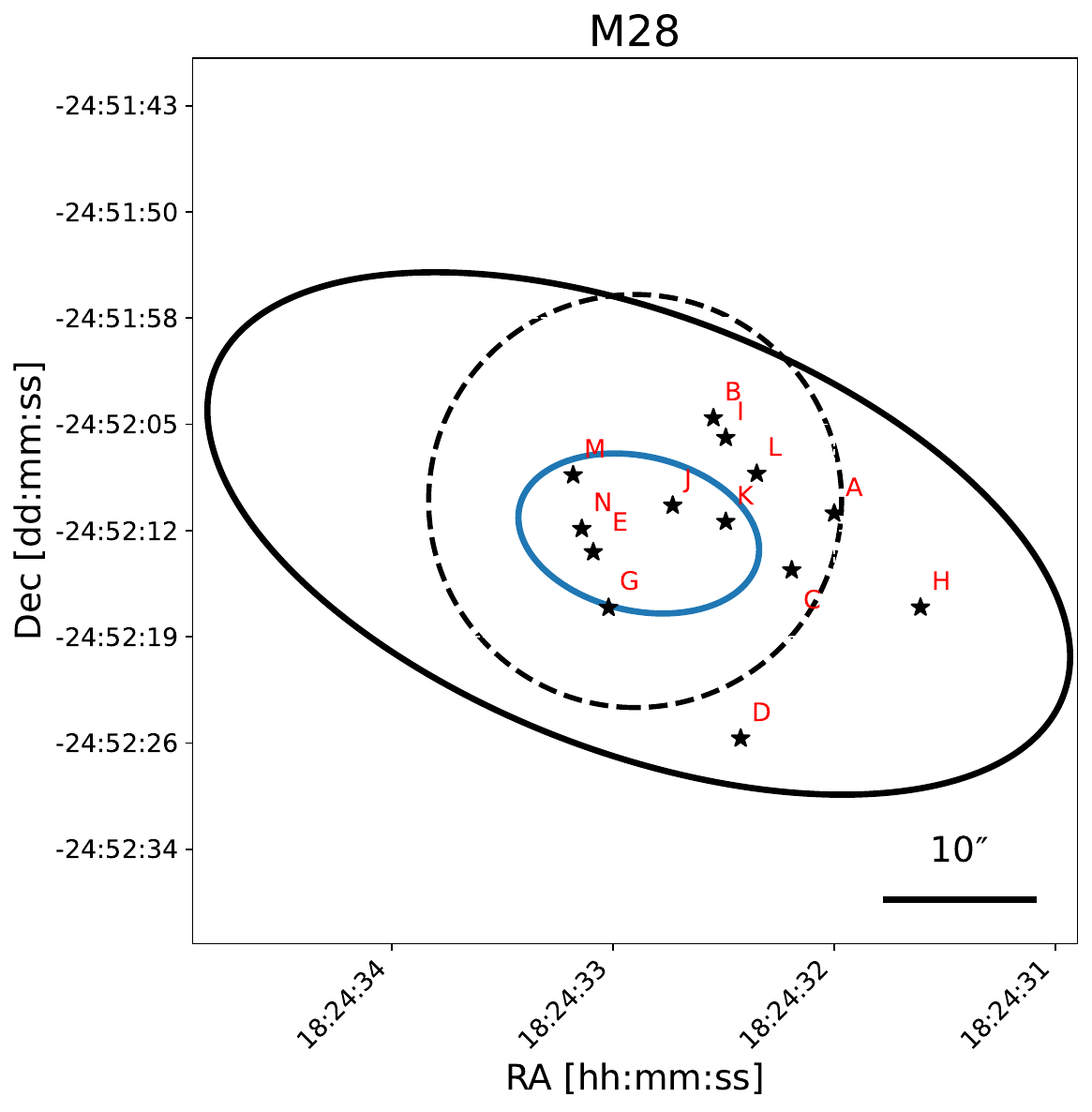}\hfill
\includegraphics[width=0.195\textwidth]{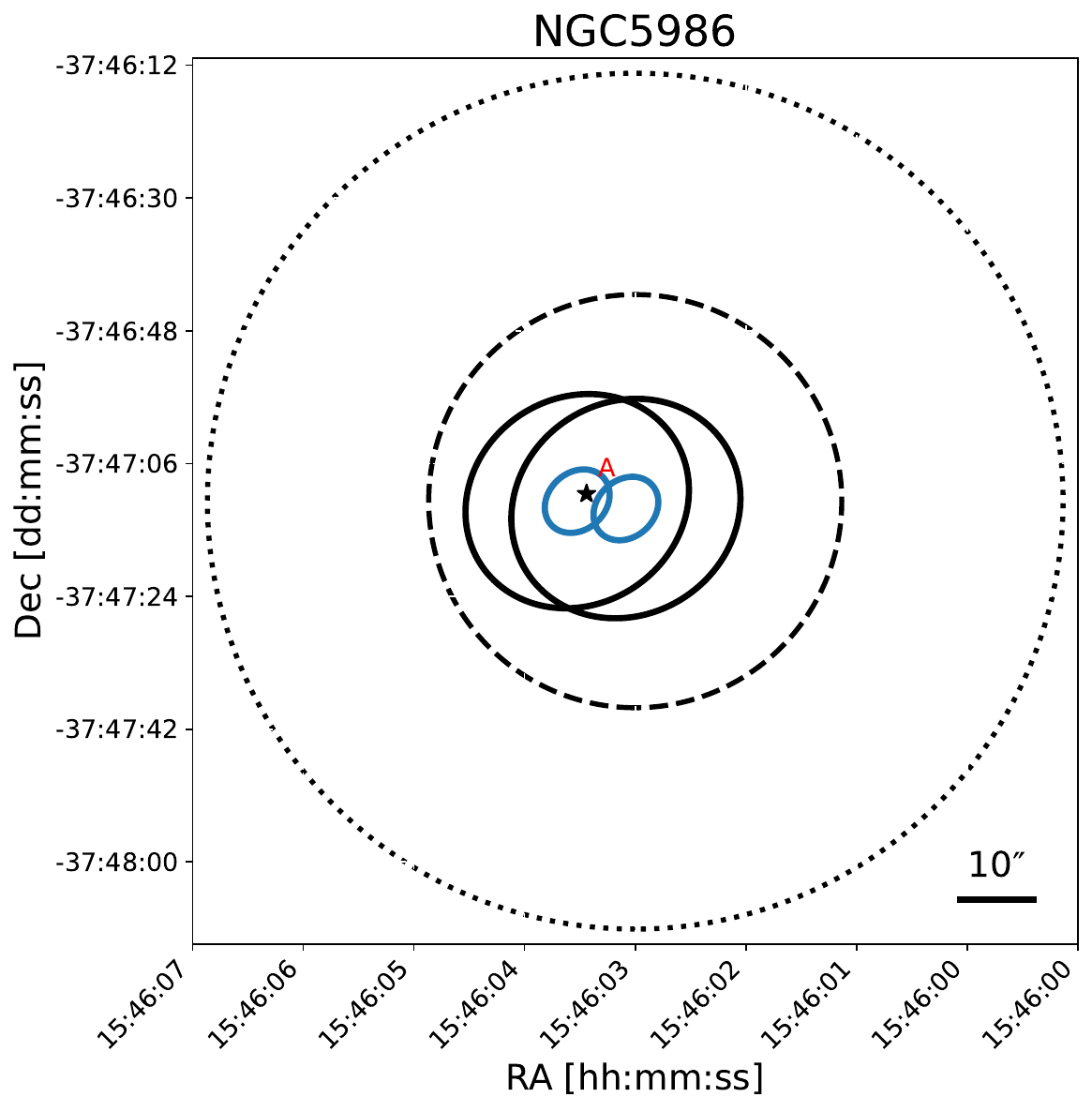}\\[-1.15em]
\includegraphics[width=0.195\textwidth]{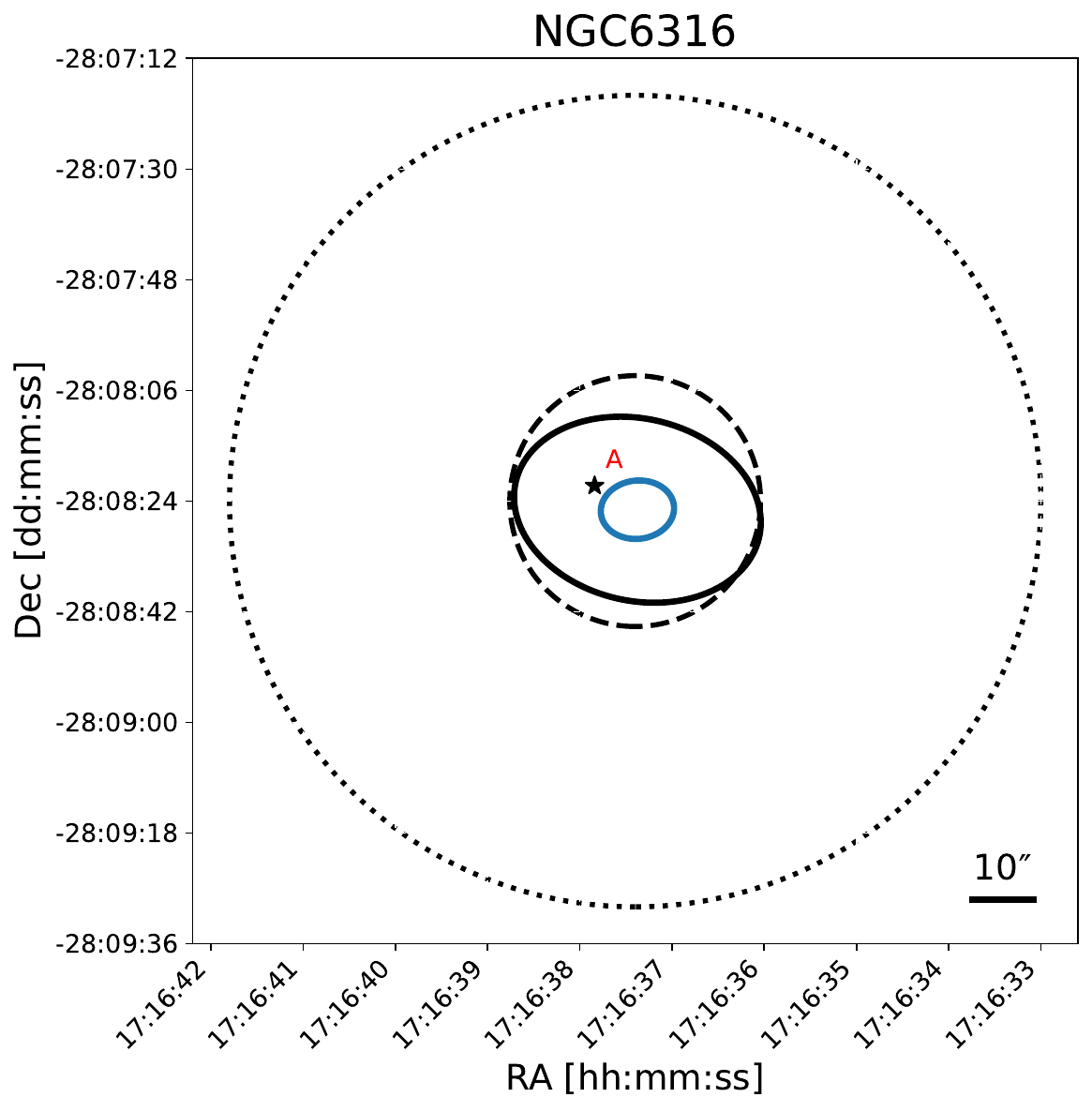}\hfill
\includegraphics[width=0.195\textwidth]{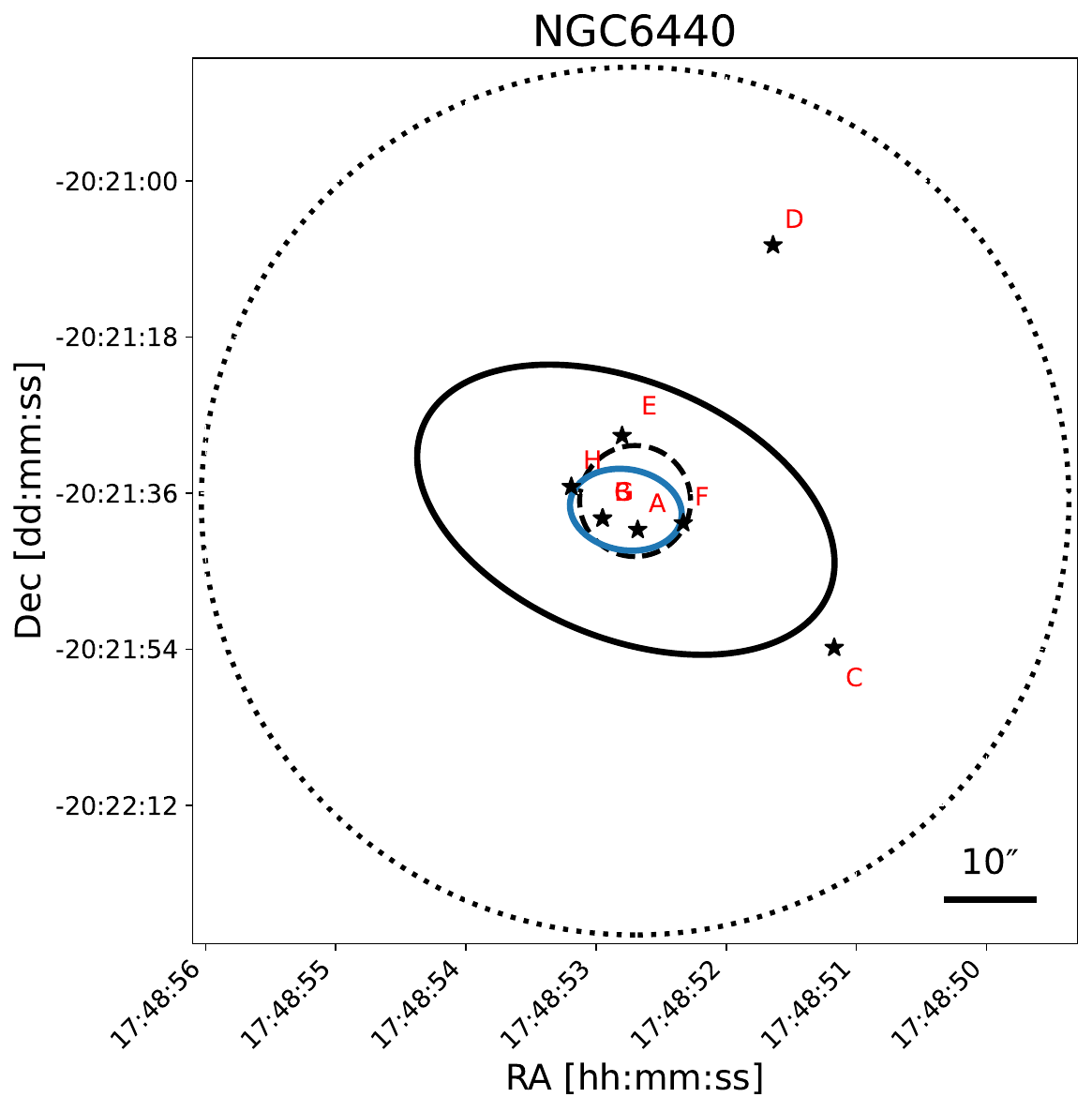}\hfill
\includegraphics[width=0.195\textwidth]{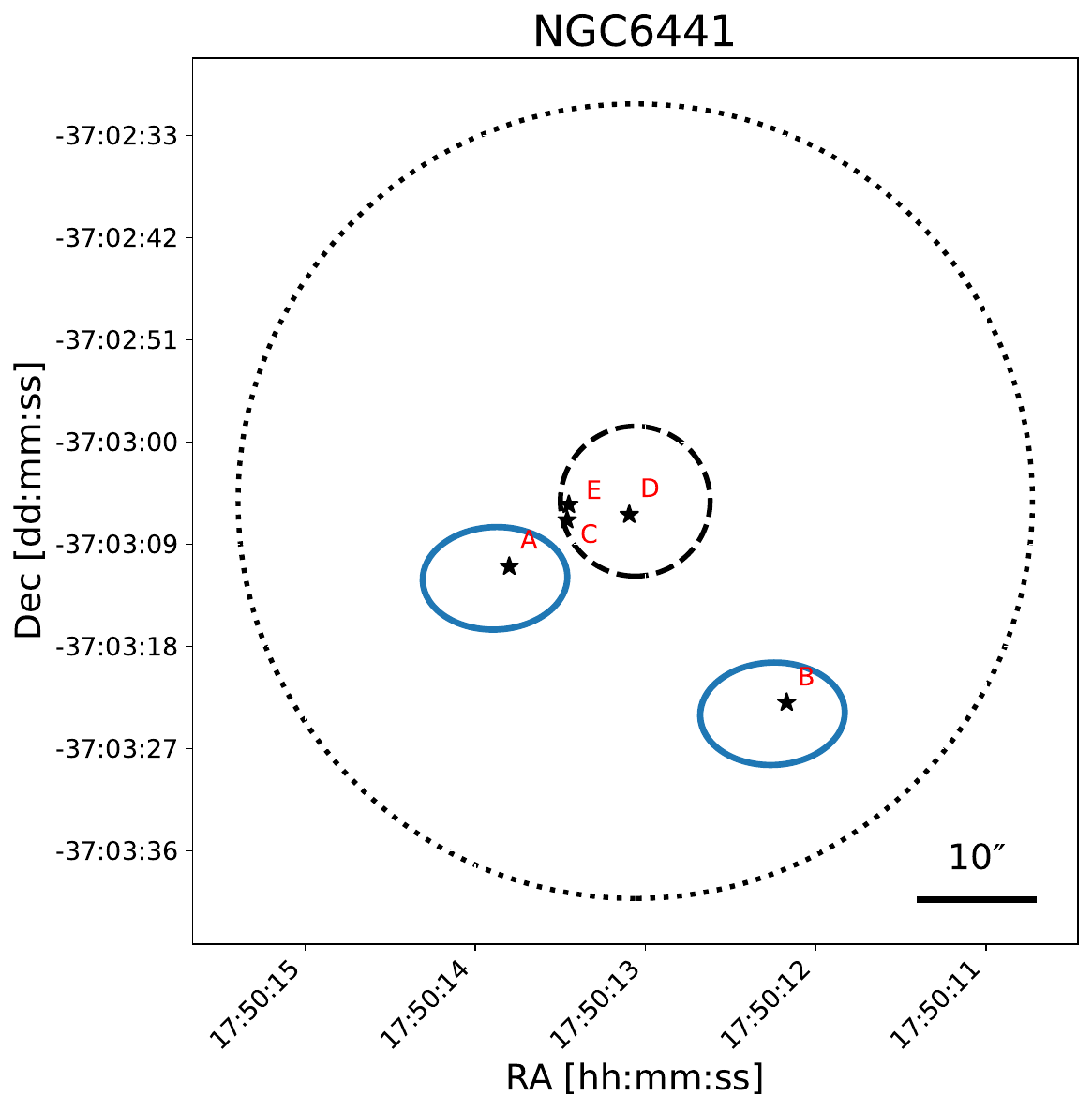}\hfill
\includegraphics[width=0.195\textwidth]{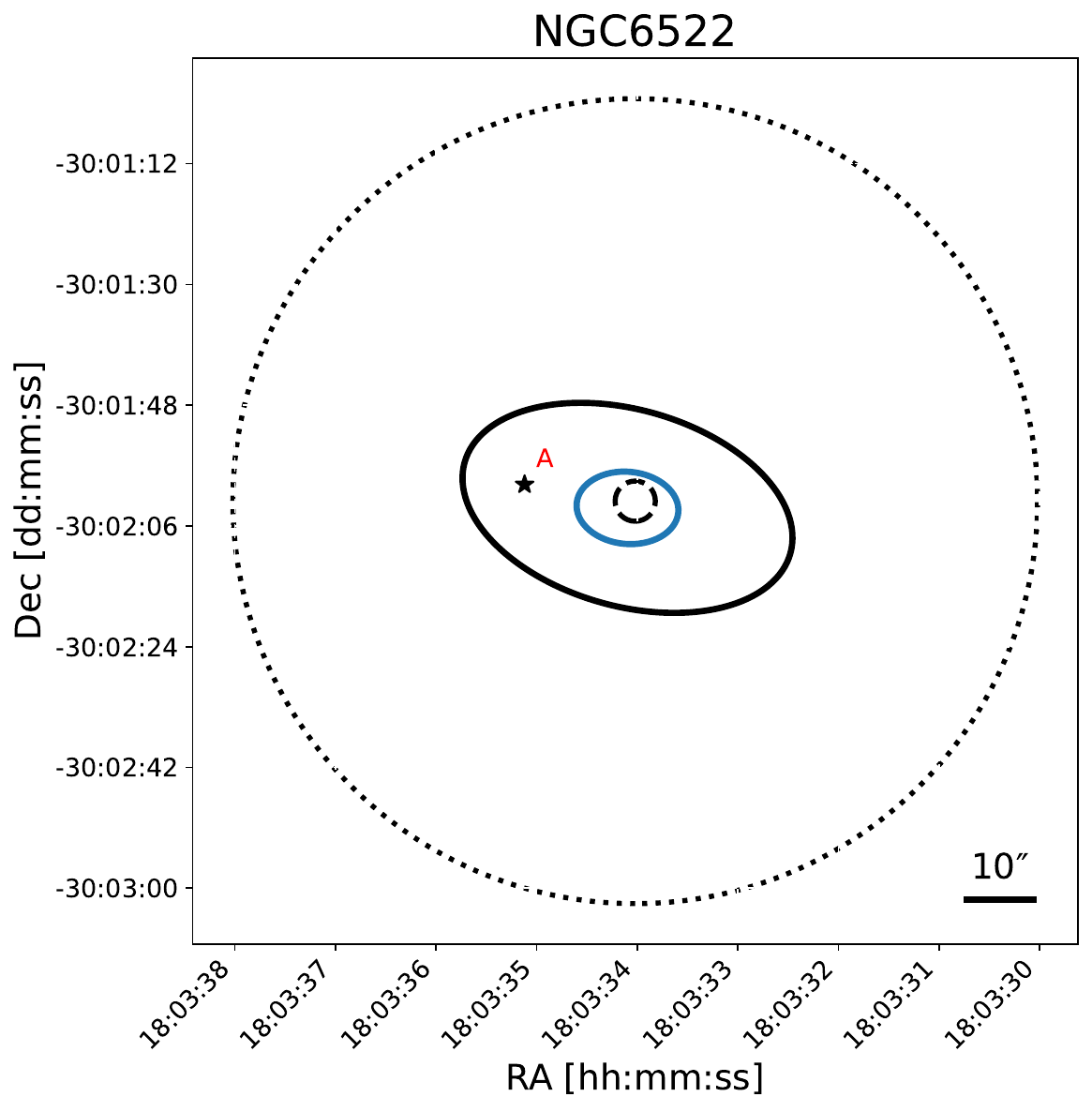}\\[-1.15em]
\includegraphics[width=0.195\textwidth]{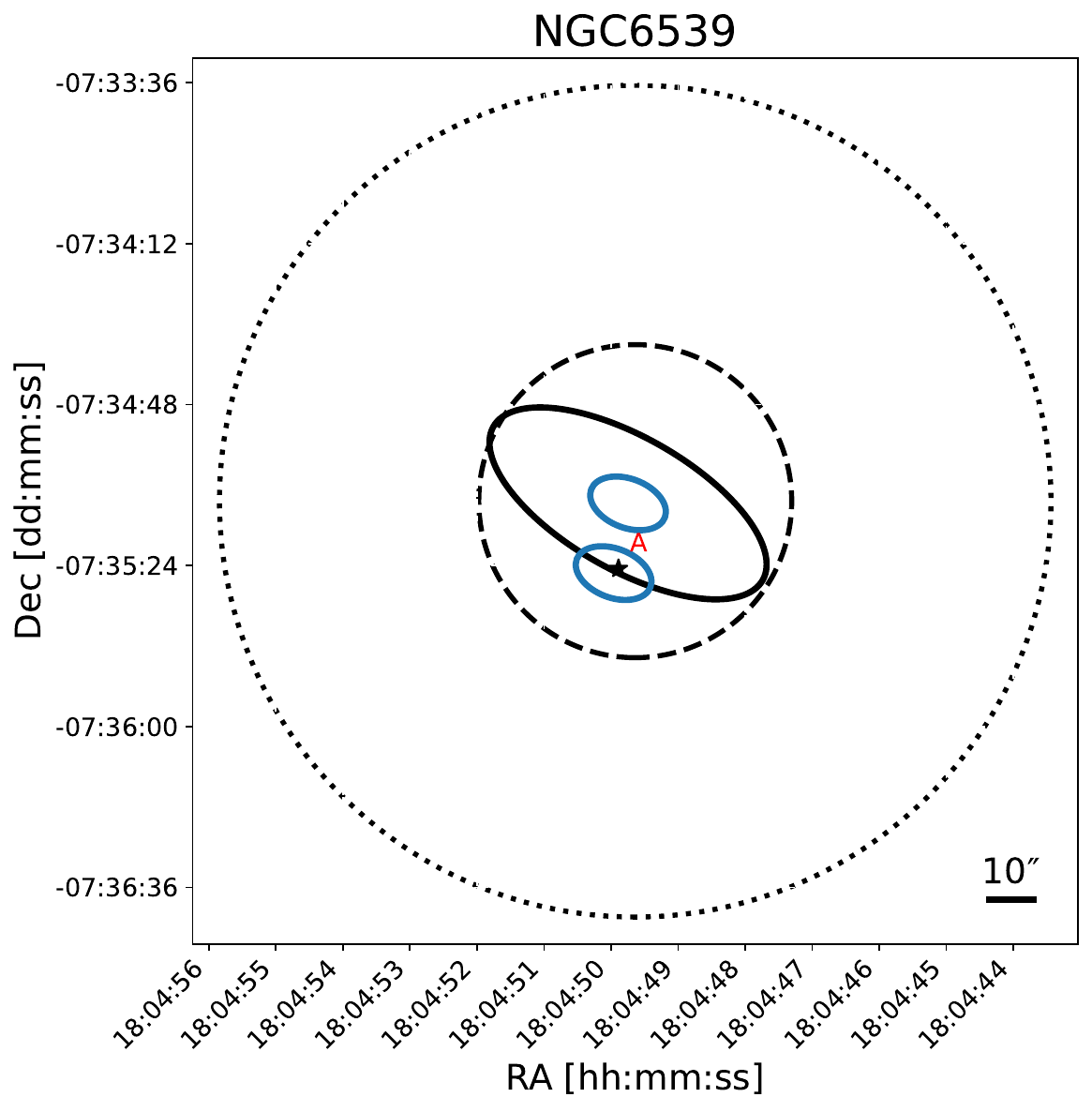}\hfill
\includegraphics[width=0.195\textwidth]{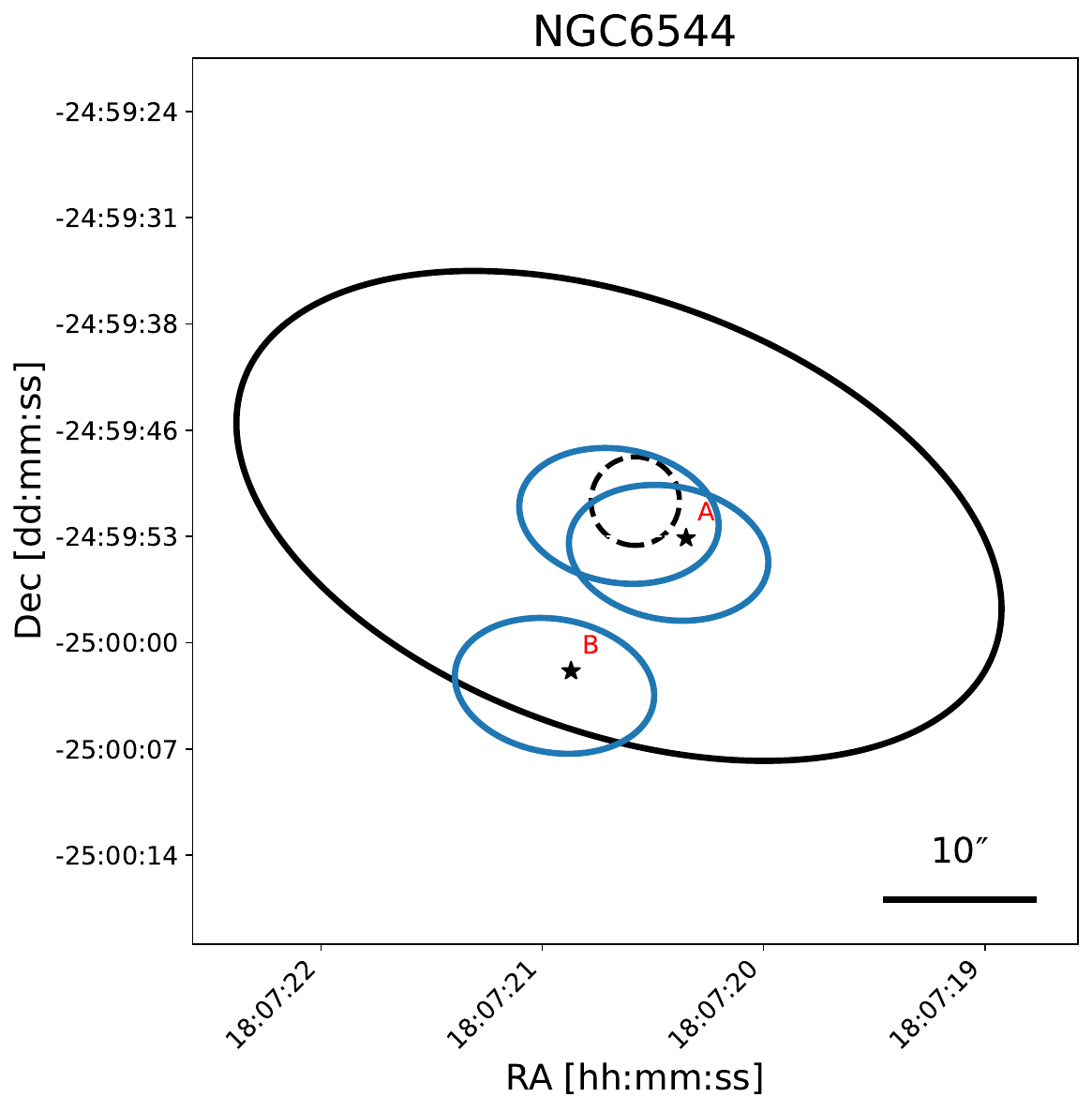}\hfill
\includegraphics[width=0.195\textwidth]{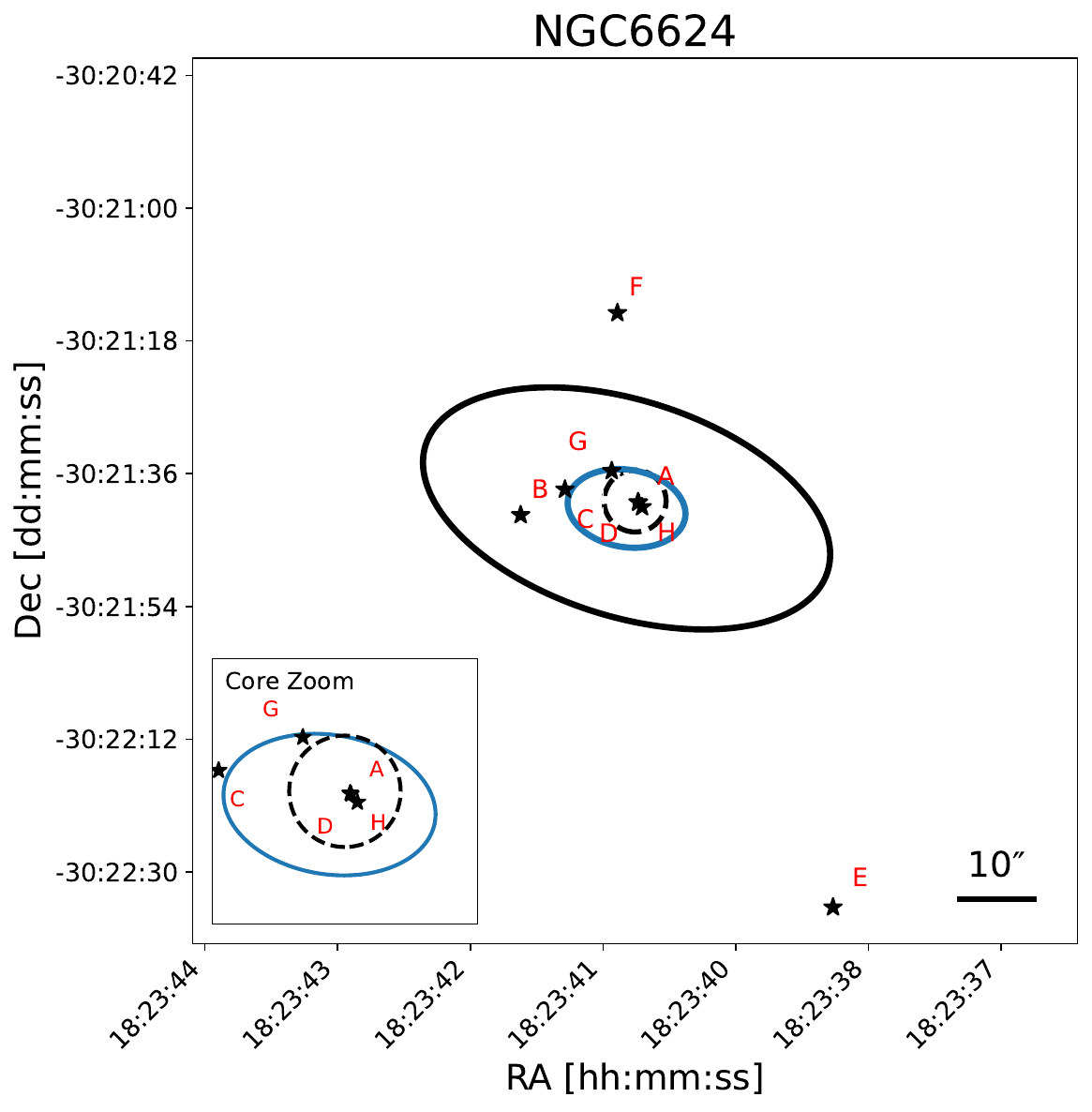}\hfill
\includegraphics[width=0.195\textwidth]{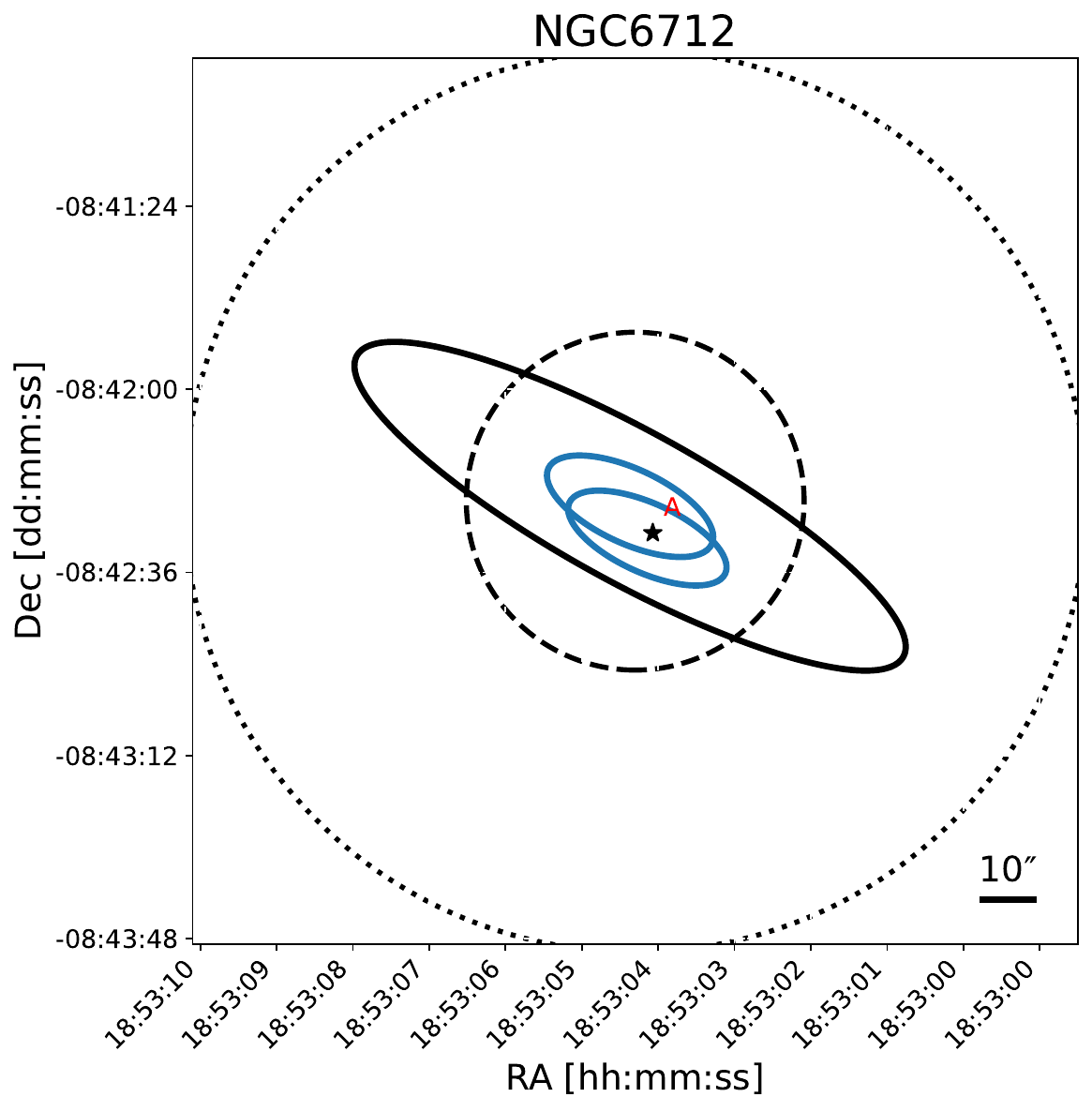}\\[-1.15em]
\includegraphics[width=0.195\textwidth]{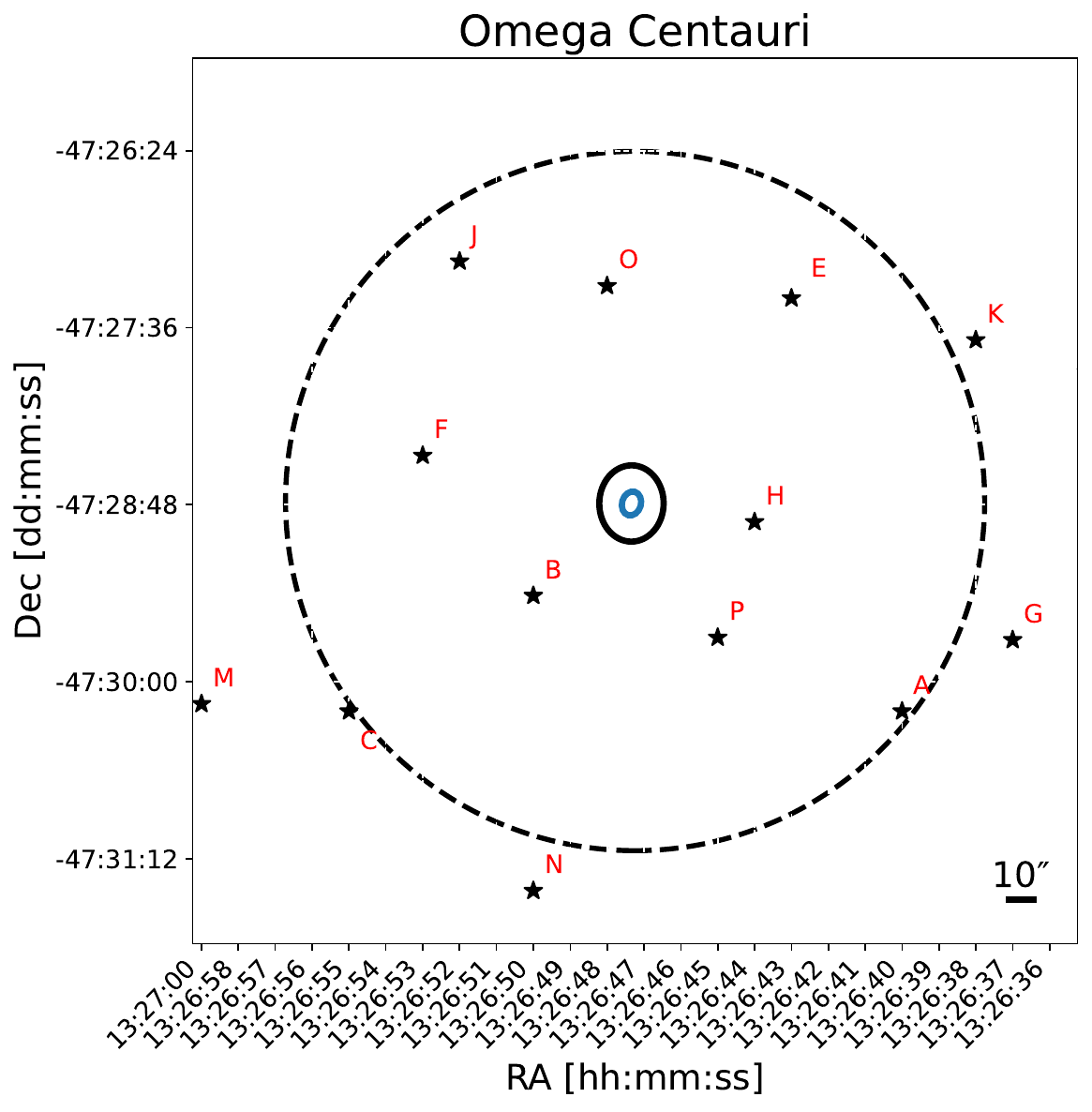}\hspace{0.08\textwidth}
\includegraphics[width=0.195\textwidth]{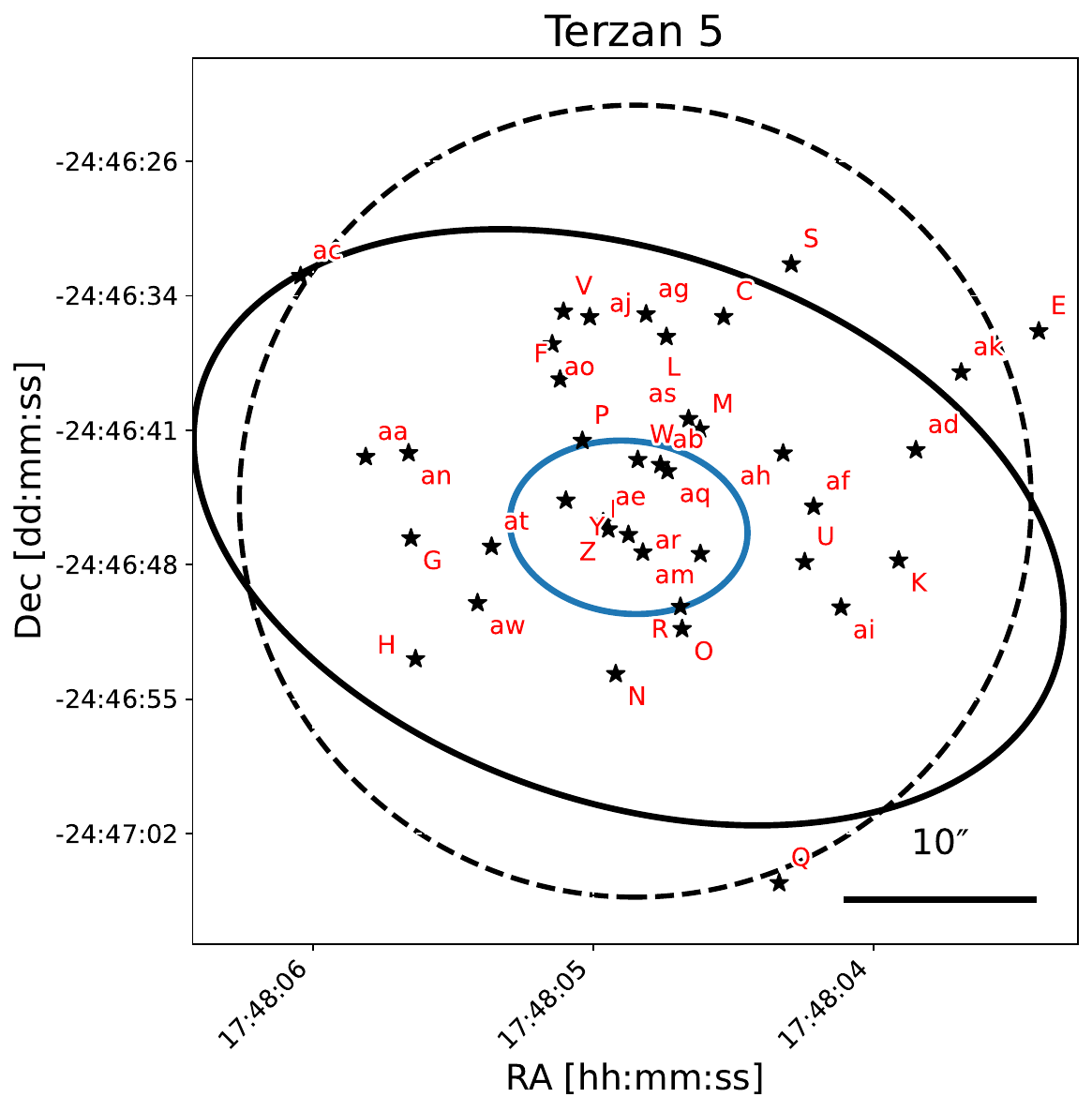}
\caption{Beam maps of the 14 GCs surveyed at S-band. Blue and black ellipses show the full-array and 1 km core tied-array beams, respectively; dashed and dotted circles mark the cluster core and half-light radii, and known pulsar timing positions are indicated where available. Cluster structural parameters were taken from the Harris GC catalogue (\url{https://physics.mcmaster.ca/~harris/mwgc.dat}), and the tied-array beam geometries were generated using the MOSAIC beam plotter (\url{https://github.com/wchenastro/Mosaic}).}
\label{fig:beammaps}
\end{figure*}

\begin{table*}[!t]
\centering
\caption{Globular cluster pointings in the first MeerKAT S-band pulsar survey.}
\label{tab:gc_observations}
\tiny
\setlength{\tabcolsep}{2.2pt}
\renewcommand{\arraystretch}{0.72}
\begin{tabular}{l l l c c c c c c}
    \toprule
    Source & RA & Dec & MJD & DM & Length & $T_{\rm samp}$ & Re‐det & New‐det \\
          & (J2000) & (J2000) &      & ($\dmunit$) & (s) & ($\mu$s) &      &        \\
    \midrule
    Glimpse‐C01 1Km  & 18:48:49.70 & $-$01:29:50.00 & 60158.752 & 491.1   & 7202  & 150 & A & C,D,E   \\
    Glimpse‐C01 FA  & 18:48:49.70 & $-$01:29:50.00 & 60158.752 & 491.1   & 7201  & 150 & A & C,F  \\

    Glimpse‐C01A 1Km & 18:48:48.21 & $-$01:29:58.27 & 60158.752 & 491.17  & 7200  & 150 &  &   \\
    Glimpse‐C01A FA & 18:48:48.21 & $-$01:29:58.27 & 60158.752 & 491.17  & 7199  & 150 &  &   \\
    \midrule
    M28 1Km          & 18:24:32.81 & $-$24:52:11.20 & 60286.630 & 119.9   & 7201  & 75  & \multicolumn{1}{c}{A,B,C,}  & $-$ \\
    & & & & & & & \multicolumn{1}{c}{E,G,H} & \\
    M28 FA          & 18:24:32.81 & $-$24:52:11.20 & 60286.630 & 119.9   & 7200  & 75  &
    \multicolumn{1}{c}{A,B,C,}  & $-$ \\
    & & & & & & & \multicolumn{1}{c}{E,G} & \\&  \\
    \midrule
    M22 1km         & 18:36:23.94 & $-$23:54:17.10 & 60138.824 &  91.2   & 7202  & 150 &  & $-$ \\
    M22 FA         & 18:36:23.94 & $-$23:54:17.10 & 60138.824 &  91.2   & 7201  & 150 &  & $-$ \\
    M22A  1km       & 18:36:25.4452 & $-$23:54:52.39 & 60138.824 &  89.107 & 7199  & 150 & A & $-$ \\
    M22B 1km         & 18:36:23.3760 & $-$23:55:20.9129 & 60138.824 &  93.772 & 7200  & 150 & B & $-$ \\
    \midrule
    NGC5986 1Km      & 15:46:03.00 & $-$37:47:11.10 & 60142.901 &  92     & 7201  & 75  &  & $-$ \\
    NGC5986 FA      & 15:46:03.00 & $-$37:47:11.10 & 60142.901 &  92     & 7198  & 75  &  & $-$ \\
    NGC5986A 1km    & 15:46:03.44 & $-$37:47:10.1 & 60142.901 &  92.17  & 7201  & 75  & A & $-$ \\
    NGC5986A FA    & 15:46:03.44 & $-$37:47:10.1 & 60142.901 &  92.17  & 7201  & 75  & A & $-$ \\
    \midrule
    NGC6316 1Km      & 17:16:37.30 & $-$28:08:24.40 & 60398.057 & 172.1   & 7202  & 75  & A & $-$ \\
    NGC6316 FA      & 17:16:37.30 & $-$28:08:24.40 & 60398.057 & 172.1   & 7199  & 75  &  & $-$ \\
    \midrule
    NGC6440 1Km      & 17:48:52.70 & $-$20:21:36.90 & 60314.496 & 222.51  & 7200  & 75  & A & $-$ \\
    NGC6440 FA      & 17:48:52.70 & $-$20:21:36.90 & 60314.496 & 222.51  & 7199  & 75  & A,B,E & $-$ \\
    \midrule
    NGC6441 FA      & 17:50:13.80 & $-$37:03:11.00 & 60296.589 & 236.6   & 7202  & 75  &  & $-$ \\
    \midrule
    NGC6522 1Km      & 18:03:34.02 & $-$30:02:02.30 & 60259.665 & 192.6   & 7200  & 75  & A & $-$ \\
    NGC6522 FA      & 18:03:34.02 & $-$30:02:02.30 & 60259.665 & 192.6   & 7199  & 75  & A & $-$ \\
    \midrule
    NGC6539 1Km      & 18:04:49.68 & $-$07:35:09.1 & 60138.693 & 186     & 7202  & 75  & A & $-$ \\
    NGC6539 FA      & 18:04:49.68 & $-$07:35:09.1 & 60138.693 & 186     & 7201  & 75  & A & $-$ \\
    NGC6539A FA    & 18:04:49.8954 & $-$07:35:24.69 & 60138.693 & 186.316 & 7200  & 75  & A & $-$ \\
    \midrule
    NGC6544 1Km     & 18:07:20.58 & $-$24:59:50.40 & 60126.971 & 135.5   & 7202  & 75  & A,B & $-$ \\
    NGC6544 FA      & 18:07:20.58 & $-$24:59:50.40 & 60126.971 & 135.5   & 7201  & 75  & A,B & $-$ \\
    NGC6544A FA     & 18:07:20.3556 & $-$24:59:52.9015 & 60126.971 & 134.004 & 7199  & 75  & A,B & $-$ \\
    NGC6544B FA     & 18:07:20.8712 & $-$25:00:01.915 & 60126.971 & 137.15  & 7200  & 75  & A,B & $-$ \\
    \midrule
    NGC6624 1Km     & 18:23:40.51 & $-$30:21:39.70 & 60329.502 &  86.2   & 7200  & 75  & A,B,D  & $-$ \\
    NGC6624 FA     & 18:23:40.51 & $-$30:21:39.70 & 60329.502 &  86.2   & 7199  & 75  & A,B,D & $-$ \\
    \midrule
    NGC6712 1Km    & 18:53:04.30 & $-$08:42:22.00 & 60144.755 & 155     & 7201  & 75  &  & $-$ \\
    NGC6712 FA      & 18:53:04.30 & $-$08:42:22.00 & 60144.755 & 155     & 7200  & 75  &  & $-$ \\
    NGC6712A FA     & 18:53:04.07409 & $-$08:42:28.254 & 60144.755 & 155.125     & 7199  & 75  & A & $-$ \\
    \midrule
    OmegaCen3 1Km   & 13:26:47.24 & $-$47:28:46.50 & 60330.964 & 100.4   & 7201  & 75  &  & $-$ \\
    OmegaCen3 1km   & 13:26:47.24 & $-$47:28:46.50 & 60331.055 & 100.4   & 7201  & 75  &  & $-$ \\
    \midrule
    Ter5 1km        & 17:48:04.80 & $-$24:46:45.00 & 60251.453 & 237.75  & 7199  & 75 & \multicolumn{1}{c}{A,C,E,G,I,L,M} & $-$ \\
& & & & & & & \multicolumn{1}{c}{N,O,V,W,Y,Z} & \\
& & & & & & & \multicolumn{1}{c}{ae,ai,am} & \\
    Ter5 FA         & 17:48:04.80 & $-$24:46:45.00 & 60251.453 & 237.75  & 7199  & 75 & \multicolumn{1}{c}{A,I,L,M,W,Z}  &  $-$\\

    \bottomrule
  \end{tabular}
\par\vspace{0.1em}
\begin{minipage}{0.97\textwidth}
\raggedright\tiny
\textbf{Notes.} For each target, we list the cluster identifier, J2000 right ascension and declination of the telescope pointing, observation date in MJD, nominal cluster DM used for de-dispersion, total integration length, and sampling interval. Columns `Re-det' and `New-det' indicate whether any known pulsar was re-detected or a new pulsar was discovered in that pointing. The DM range searched for each GC was $\pm20\%$ of the nominal cluster DM (the median for clusters with multiple pulsars and the DM of pulsar A in clusters with only one known pulsar), with DM step sizes calculated using \textsc{PRESTO}'s \texttt{DDplan.py}.
\end{minipage}
\end{table*}
\FloatBarrier

\section{Flux density estimation from folded pulse profiles}
\label{flux_density_method}
\begingroup
\setlength{\abovedisplayskip}{3pt}
\setlength{\belowdisplayskip}{3pt}
\setlength{\abovedisplayshortskip}{2pt}
\setlength{\belowdisplayshortskip}{2pt}

We derive here the expression used to estimate the phase-averaged pulsar flux density from folded pulse profiles. For an observation with integration time $t_{\rm obs}$ and usable bandwidth $\Delta\nu$, the rms noise of the dedispersed time series is given by the radiometer equation in terms of the effective system-equivalent flux density (SEFD):
\begin{equation}
\sigma_{\rm ts} =
\frac{\mathrm{SEFD}}
{\sqrt{n_{\rm pol}\,\Delta\nu\,t_{\rm obs}}},
\end{equation}
where $n_{\rm pol}$ is the number of summed orthogonal polarisations.

After folding the time series into $N_{\rm bin}$ pulse-phase bins, each bin contains an effective integration time of $t_{\rm obs}/N_{\rm bin}$. The rms noise per phase bin is therefore
\begin{equation}
\sigma_{\rm bin} =
\frac{\mathrm{SEFD}}
{\sqrt{n_{\rm pol}\,\Delta\nu\,(t_{\rm obs}/N_{\rm bin})}}
=
\frac{\mathrm{SEFD}}
{\sqrt{n_{\rm pol}\,\Delta\nu\,t_{\rm obs}}}
\sqrt{N_{\rm bin}}.
\end{equation}

Let $N_{\rm on}$ denote the number of phase bins containing pulsed emission and $N_{\rm off} = N_{\rm bin} - N_{\rm on}$ the number of off-pulse bins used to determine the baseline level. The integrated pulse amplitude above the baseline is
\begin{equation}
A = \sum_{i=1}^{N_{\rm on}} (p_i - b),
\end{equation}
where $p_i$ are the on-pulse bin values and $b$ is the mean off-pulse baseline.

The uncertainty in the integrated pulse amplitude arises from both the noise in the on-pulse bins and the uncertainty in the baseline estimate derived from the off-pulse region. Propagating these contributions gives
\begin{equation}
\sigma_A =
\sigma_{\rm bin}
\sqrt{
N_{\rm on}
\left(
1 + \frac{N_{\rm on}}{N_{\rm off}}
\right)
}.
\end{equation}
The signal-to-noise ratio of the folded profile is therefore
\begin{equation}
{\rm S/N} = \frac{A}{\sigma_A}.
\end{equation}

The phase-averaged flux density $S$ satisfies
\begin{equation}
A = S\,N_{\rm bin}.
\end{equation}
Substituting for $A$ and $\sigma_A$ and rearranging yields
\begin{equation}
S =
\frac{{\rm S/N}\times\mathrm{SEFD}}
{\sqrt{n_{\rm pol}\,\Delta\nu\,t_{\rm obs}}}
\sqrt{\frac{N_{\rm on}}{N_{\rm off}}},
\end{equation}
which is the expression used to compute the flux densities reported in this work.

\section{Effective SEFD and gain determination}
\label{app:sefd}

To convert folded pulse profile signal-to-noise ratios into phase-averaged flux densities, we require an effective system sensitivity appropriate for the coherently beamformed MeerKAT S-band observations. In the pulsar radiometer equation, this sensitivity enters through the factor $(T_{\rm rec}+T_{\rm sky})/G$, which is equivalent to the SEFD. In this appendix, we describe how this quantity is determined for our observations.

Published MeerKAT S-band performance data provide the mean SEFD as a function of frequency for a single antenna across the full receiver band. The SEFD curves used here were obtained from the official SARAO S-band capability documentation. These curves represent the intrinsic per-antenna sensitivity of the system and include receiver temperature, sky contribution, and antenna gain, but do not incorporate backend or processing losses. Because no tabulated values are provided, the SEFD curves were digitised from the published performance plots and interpolated across the observing band.

To account for the frequency dependence of the receiver response across the wide S-band bandwidth, we compute a band-averaged effective SEFD. Since signal-to-noise adds in quadrature across independent frequency channels, the effective per-antenna SEFD is evaluated using an inverse-square weighted average,
\begin{equation}
\mathrm{SEFD}_{\rm eff,ant}
=
\left(
\frac{\sum_i w_i\,\mathrm{SEFD}(f_i)^{-2}}
     {\sum_i w_i}
\right)^{-1/2},
\end{equation}
where $\mathrm{SEFD}(f_i)$ is the per-antenna SEFD at frequency $f_i$ and $w_i$ are per-channel weights. Given the low radio-frequency interference occupancy of the S-band data used in this work (typically $\gtrsim 98\%$ usable bandwidth), uniform weights are assumed.

The per-antenna effective SEFD is then scaled to the coherently beamformed tied-array beam by dividing by the number of antennas contributing to the coherent sum,
\begin{equation}
\mathrm{SEFD}_{\rm eff,array}
=
\frac{\mathrm{SEFD}_{\rm eff,ant}}{N_{\rm ant}}.
\end{equation}
For observations using the full MeerKAT array, we adopt $N_{\rm ant}$ between 58 and 64, consistent with standard S-band sensitivity assumptions. For beams formed using only the inner core of the array (baselines $\lesssim 1$~km), we adopt $N_{\rm ant}$ between 35 and 48. This procedure yields effective tied-array SEFD values of approximately $7.4$--$8.1$~Jy for the full array and $9.8$--$13.5$~Jy for the core configuration.

We note that this determination of the effective SEFD corresponds solely to the intrinsic telescope sensitivity and does not include losses associated with digitisation, detection, or backend processing, which are commonly parameterised by a multiplicative factor $\beta$ in the pulsar radiometer equation. Because signal-to-noise ratios are measured directly from the folded pulse profiles produced by the full processing chain, and no independent characterisation of backend efficiency is available, we set $\beta=1$ and do not apply an additional correction. Any residual processing losses are therefore implicitly absorbed into the measured signal-to-noise ratios. Derived flux densities scale linearly with any alternative choice of $\beta$.

\par\medskip
\begin{center}
\includegraphics[width=0.55\linewidth]{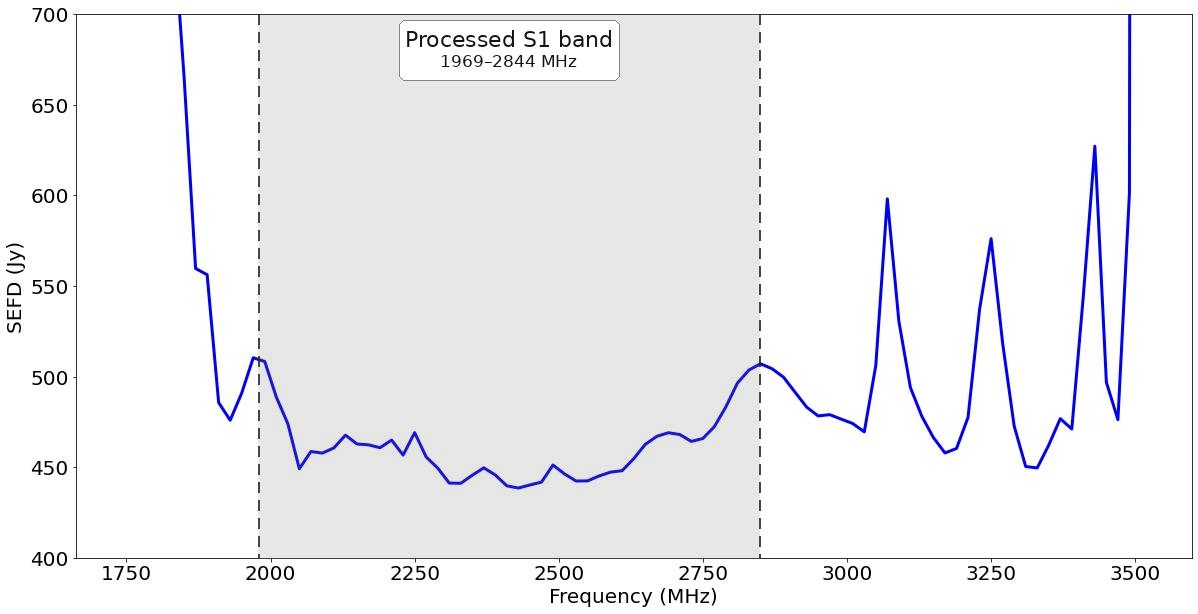}
\captionof{figure}{Mean per-antenna SEFD across the full MeerKAT S-band receiver range. The shaded region indicates the approximate frequency interval used for the S1 observations analysed in this work, corresponding to an 875\,MHz processed bandwidth centred at 2406.25\,MHz.}
\label{fig:sefd_band}
\end{center}

\endgroup

\end{appendix}

\end{document}